\documentclass[10pt,letterpaper]{article}
\usepackage[top=0.85in,left=2.75in,footskip=0.75in]{geometry}

\usepackage{amsmath,amssymb}

\usepackage{changepage}

\usepackage{textcomp,marvosym}

\usepackage{cite}

\usepackage{nameref}
\usepackage{hyperref}

\usepackage[right]{lineno}

\usepackage[nopatch=eqnum]{microtype}
\DisableLigatures[f]{encoding = *, family = * }

\usepackage[table]{xcolor}

\usepackage{array}
\usepackage{spotcolor}

\newcolumntype{+}{!{\vrule width 2pt}}

\newlength\savedwidth

\raggedright
\usepackage[aboveskip=1pt,labelfont=bf,labelsep=period,justification=raggedright,singlelinecheck=off]{caption}

\makeatletter
\renewcommand{\@biblabel}[1]{\quad#1.}
\makeatother

\usepackage{lastpage,fancyhdr,graphicx}
\usepackage{epstopdf}
\fancyheadoffset[L]{2.25in}
\fancyfootoffset[L]{2.25in}
\NewSpotColorSpace{PANTONE}
\AddSpotColor{PANTONE} {PANTONE3015C} {PANTONE\SpotSpace 3015\SpotSpace C} {1 0.3 0 0.2}
\SetPageColorSpace{PANTONE}
\definecolor{accessblue}{cmyk}{1, 0.3, 0, 0.2}
\definecolor{greycolor}{cmyk}{0,0,0,.8}

\usepackage{amsfonts}
\usepackage{bm}
\usepackage{subcaption}
\usepackage[export]{adjustbox}
\usepackage{tabularx}
\usepackage{booktabs}
\usepackage{multirow}
\usepackage{makecell}
\usepackage{algorithmic}
\usepackage{comment}
\usepackage{ragged2e}
\usepackage{wrapfig}
\usepackage{url}
\usepackage{rotating}
\usepackage{ifthen}
\usepackage{pifont}
\usepackage{tikz}
\usepackage{pgfplots}
\usepackage{placeins}

\pgfplotsset{compat=1.18}
\usetikzlibrary{calc,decorations.pathmorphing,patterns,shapes,shadows,decorations.pathreplacing,arrows,positioning,arrows.meta,snakes,shapes.symbols}
\usepgfplotslibrary{groupplots}

\begin{document}
\vspace*{0.2in}

\begin{flushleft}
\textbf{Ideological discrepancy between Facebook pages and shared news is associated with reaction homogeneity}
\newline

Thiago Magrin\textsuperscript{1},
Jordan Kobellarz\textsuperscript{1},
Pedro O.S. Vaz-de-Melo\textsuperscript{2},
Thiago H. Silva\textsuperscript{1,3*}
\\
\bigskip
\textbf{1} Universidade Tecnológica Federal do Paraná, Curitiba, Brazil
\\
\textbf{2} Universidade Federal de Minas Gerais, Belo Horizonte, Brazil
\\
\textbf{3} University of Toronto, Toronto, Canada
\\
\bigskip

* th.silva@utoronto.ca

\end{flushleft}

\section*{Abstract}

Cross-cutting political content is often expected to mitigate polarization, yet encounters across ideological lines do not necessarily foster more constructive engagement. We analyzed 4,584 posts from 1,243 Facebook pages sharing 761 political news links in a dataset centered on Brazil's 2018 presidential election, and introduced Bias Discrepancy, which measures how far a shared link lies from the page's ideological sharing repertoire. We examined how Bias Discrepancy related to Content Reaction Positivity (how positively the same link was received in other posts), the Reaction Homogeneity Index (whether a post's own reactions were one-sided or mixed), and the toxicity of the page's accompanying text. Reaction homogeneity was lowest at both extremes of Bias Discrepancy, and posts furthest from the page's repertoire had the highest average toxicity. In a multivariable model, greater Bias Discrepancy was associated with lower reaction homogeneity, with a magnitude similar to the link's content bias and toxicity and smaller than Content Reaction Positivity. When a page shared news that departed from its repertoire, its accompanying text tended to be more toxic and its audience's reactions more mixed.


\section*{Introduction}

The rise of social media as a primary source of political information has transformed news consumption and how individuals engage with political content \cite{levy2021social}. On Facebook, both ranking algorithms and individual choices shape exposure to political content. In a field experiment that randomized subscriptions to news outlets, Facebook's ranking supplied fewer posts from counter-attitudinal outlets even to users who had subscribed to them \cite{levy2021social}, while across the platform as a whole individual choices did more than ranking to limit cross-cutting exposure \cite{bakshy2015exposure}. By curating personalized content based on users' past interactions, platforms may in turn contribute to the formation of ``filter bubbles,'' informational ecosystems in which exposure to divergent views is limited and pre-existing beliefs are reinforced \cite{cinelli2021echo, pariser2011filter}. This process has been associated with increased polarization, as homophilous networks \cite{mcpherson2001birds} promote the formation of ideologically homogeneous groups where misinformation can spread quickly \cite{del2016spreading}.

These structural conditions also have affective consequences. Because echo chambers amplify political beliefs while limiting exposure to opposing viewpoints \cite{cinelli2021echo, vosoughi2018spread}, hostility between opposing groups may intensify \cite{bail2018exposure}, and interactions may become increasingly toxic \cite{Kobellarz2024}. Elections are one setting where these dynamics play out, since social media is used for voter mobilization and for building campaign narratives. Brazil's 2018 presidential election is one example. In that election, Jair Bolsonaro built his campaign around an anti-corruption agenda and used social media as his main channel to communicate with voters \cite{de2023visual}, and retweet networks during the election showed tight clustering around ideological poles \cite{Kobellarz2019}. These dynamics reflected an intensification of affective polarization, more pronounced toward candidates than toward parties, and accompanied the growth and radicalization of the political right \cite{fuks2022polarizaccao}. Later analyses of the 2022 election indicate that polarization continued to increase \cite{cunha2025caracterizando}.

Most existing work on these dynamics focuses on network exposure~\cite{bakshy2015exposure, guess2021exposure}, content diffusion~\cite{vosoughi2018spread, del2016spreading}, or structural
polarization metrics~\cite{garimella2018quantifying, conover2011political}.
By comparison, less attention has been given to what happens when a Facebook page shares
political news that departs from its ideological sharing repertoire. Such departures matter because cross-cutting content does not necessarily produce
more constructive engagement, and prior work shows that encounters across
ideological boundaries can be accompanied by negative affect and
nonconstructive interaction \cite{bail2018exposure, cetinkaya2025cross}.
Studying them at the level of individual Facebook posts makes it possible to
connect ideological discrepancy to the reactions and language of a
particular post.

We address this gap using Brazil's 2018 presidential election as a case study and ask two research questions.
(RQ1) How is the ideological discrepancy between a shared news link and a page's ideological sharing repertoire related to audience reactions and to the toxicity of the accompanying post text? (RQ2) How does the association between this ideological discrepancy and the homogeneity of audience reactions compare with associations involving other page-, content-, and post-level characteristics?

The posts we analyze come from public Facebook pages. We call these accounts \textit{pages} throughout, rather than publishers, because most of them are not news organizations. They are public Facebook pages and verified accounts, and the term keeps them distinct from the \textit{audience} whose reactions we measure. For each page, we characterize its ideological sharing repertoire with a bias score ($b_p$) estimated from the ideologically labeled news links it has shared, and define Bias Discrepancy ($\Delta_b$) as the absolute difference between this score and the bias score of a particular shared link ($b_c$). A high $\Delta_b$ identifies a link that departs from that repertoire. It does not, however, establish the page's independently held ideology or its reason for sharing that link. Bias Discrepancy is the measure we use throughout for the ideological discrepancy in the two research questions, and we relate it to how audiences respond and to the language pages use when sharing. Audience response is captured by two measures: Content Reaction Positivity ($CRP$), the positivity of reactions to other posts sharing the same link, and the Reaction Homogeneity Index ($RHI$), which records whether a post's reactions fall predominantly on one side or are divided between them. Toxicity ($\text{Tox}$), by contrast, characterizes the post rather than the audience, quantifying hostile language in the text written by the page alongside the link. The linked articles are further characterized through zero-shot topic classification. Finally, to address RQ2, we model $RHI$ using a Generalized Linear Mixed Model (GLMM) with a Beta distribution, estimating how Bias Discrepancy, Content Reaction Positivity, toxicity, and content topics relate to reaction homogeneity while accounting for differences across pages and shared news links.

This framework yields two main contributions. First, we connect departures from a Facebook page's ideological sharing repertoire to audience reactions and to the toxicity of the accompanying post text. We found that reaction homogeneity was lowest at both ends of the Bias Discrepancy distribution, that posts in the highest Bias Discrepancy quartile had the highest average toxicity, and that Content Reaction Positivity was lowest in the lowest quartile. Second, we use a multivariable mixed model to compare how Bias Discrepancy, Content Reaction Positivity, toxicity, and other predictors relate to reaction homogeneity. In that model, Content Reaction Positivity had the strongest association, followed at similar magnitudes by Bias Discrepancy, content bias, and toxicity, and the negative association between Bias Discrepancy and reaction homogeneity persisted across the sensitivity analyses.

The remainder of this paper is structured as follows: Section `Related work' reviews the relevant literature. Section `Problem Statement' defines the research problem. Section `Methodology' describes the dataset and analytical methods. Section `Results' presents the empirical findings, and Section `Conclusion' summarizes the main findings, their implications, and limitations. Finally, Section `Ethical considerations' addresses the ethical and privacy aspects of the study.

\section*{Related work}

Research in computational social science has examined how user behavior, social ties, and algorithmic systems contribute to ideological echo chambers \cite{flaxman2016filter,cinelli2021echo} and political homophily \cite{mcpherson2001birds,colleoni2014echo,boutyline2017echo}, across multiple platforms and methods. More recent work reports that social media platforms shape perceptions of polarization \cite{Overgaard_2024}, that social sorting reinforces affective polarization across digital interactions \cite{mason2025social}, and that partisanship can either bridge or deepen online echo chambers \cite{Erickson_2023}. Within this literature, three lines of research inform the present study: (i) how exposure to cross-cutting content is shaped and what follows from it, (ii) how existing polarization metrics capture ideological divisions in online networks, and (iii) how affective responses and emotional reactions structure engagement with political content.

Research on algorithmic curation examines how recommendation systems shape users' exposure to ideologically diverse content, and how much of that shaping is algorithmic rather than chosen. For example, \cite{bakshy2015exposure} found that individual choices, more than algorithmic ranking, limit cross-cutting content on Facebook, while \cite{guess2021exposure} showed that exposure to untrustworthy websites during the 2016 U.S. election was concentrated among a subset of users with strong preferences for pro-attitudinal information. Beyond Facebook, personalization on YouTube has been associated with the reinforcement of echo chambers \cite{ribeiro2020auditing}. Separately, a Twitter field experiment found that exposure to opposing political views can increase polarization among some users \cite{bail2018exposure}. By contrast, a field experiment on Facebook found the opposite direction for affective attitudes, with exposure to counter-attitudinal outlets reducing negative feeling toward the opposing party \cite{levy2021social}. Studies of cross-partisan interaction add a further qualification. While \cite{cetinkaya2025cross} observed significant engagement among users with differing political views on Twitter, these interactions were often characterized by negative, nonconstructive stances, suggesting that breaking echo chambers does not necessarily lead to constructive dialogue. Similarly, examining downstream effects on Reddit, \cite{xia2025integrated} showed that receiving a cross-party reply rarely increases subsequent engagement with out-party communities, instead reinforcing in-party participation. These findings suggest that exposure patterns alone do not explain how polarization appears in online interactions.

A second line of research focuses on measuring polarization in online networks. Metrics such as $RP(H)$ \cite{kobellarz2022reaching} and controversy frameworks \cite{garimella2018quantifying,matakos2017measuring} have been used to quantify ideological divisions in retweet networks. These approaches describe the structural organization of polarized communication, particularly in diffusion-based platforms such as Twitter. However, they primarily capture patterns of information flow and network structure, making them less suited to measuring engagement dynamics on platforms like Facebook, where interaction often occurs through reactions and comments rather than content resharing. For this reason, we use an $RP(H)$-derived score only to characterize news links, and we measure audience response on Facebook through reactions.

A third line of research examines the role of affective dynamics in online political communication. Prior work links textual features to affective polarization, examining how sentiment, emotion, and misinformation structure political discourse online. For instance, political alignment can often be inferred directly from language use and framing \cite{conover2011predicting}, and false news has been found to spread farther and faster than true news, with responses to false stories more often expressing fear, disgust, and surprise \cite{vosoughi2018spread}. Platform-specific emotional dynamics further influence engagement patterns. For example, sadness has been shown to play a prominent role in the sharing of political content on Facebook \cite{de2021sadness}, where content has also been found to spread mainly within homogeneous clusters of users \cite{del2016spreading}. These affective asymmetries may also be amplified by algorithmic curation mechanisms that promote emotionally engaging content \cite{guess2023}. This matters for the present study because, on platforms such as Facebook, emotional responses are expressed directly through reaction mechanisms, which provide observable signals of collective audience behavior.

These three lines of research describe how ideological boundaries shape exposure and diffusion, and how cross-cutting encounters can be accompanied by negative affect. They say less, however, about a single sharing event, in which a page shares a link that departs from its ideological sharing repertoire. Two things can be observed at that level. Reactions show how positively a link is received across posts and how homogeneous the response to a particular post is, and the toxicity of the accompanying text shows the language the page itself uses to frame the link. We do not treat reaction homogeneity or toxicity as direct measures of political polarization.

\section*{Problem Statement}
\label{sec:problem}

We consider a set of political news links, each with a content bias score $b_c \in [-1,+1]$ ranging from left- to right-leaning, and the Facebook pages that share them. Each page $p$ has a page bias $b_p \in [-1,+1]$, computed from the left-, center-, or right-leaning classification of the links in all of its collected posts, as described in the Political bias metrics subsection (Eq.~\ref{eqBc}). A link the page shared more than once therefore counts once per post. Page bias summarizes the page's ideological sharing repertoire, with each link weighted by the number of times the page shared it. It characterizes the page's observed sharing behavior rather than an independently established or time-invariant political ideology.

For each Facebook post $i$, in which page $p$ shares news link $c$, we
define Bias Discrepancy $\Delta_{b,i}$ as the absolute difference between $b_c$ and $b_p$:
\begin{equation}
    \Delta_{b,i}=|b_c-b_p|.
    \label{eq:deltabias}
\end{equation}
Larger values thus indicate that the shared news link lies farther from the page's
ideological sharing repertoire.

Each post $i$ is further described by three measures defined in the Methodology. Content Reaction Positivity ($CRP_i$) is the positivity of the reactions received by other posts sharing link $c$. The Reaction Homogeneity Index ($RHI_i$) measures whether a post's reactions are predominantly positive or negative, or more evenly distributed between these two categories. Toxicity ($\text{Tox}_i$) quantifies hostile language in the text that page $p$ wrote alongside the link. The problem has two parts, corresponding to the research questions. The first is to describe how $CRP_i$, $RHI_i$, and $\text{Tox}_i$ vary with $\Delta_{b,i}$ (RQ1). The second is to estimate the association between $\Delta_{b,i}$ and $RHI_i$ while accounting for other page-, content-, and post-level characteristics, and to compare its magnitude with theirs (RQ2).

\section*{Methodology}
\label{cap:methodology}

\begin{figure}[htb]
\centering
\includegraphics[width=.7\linewidth]{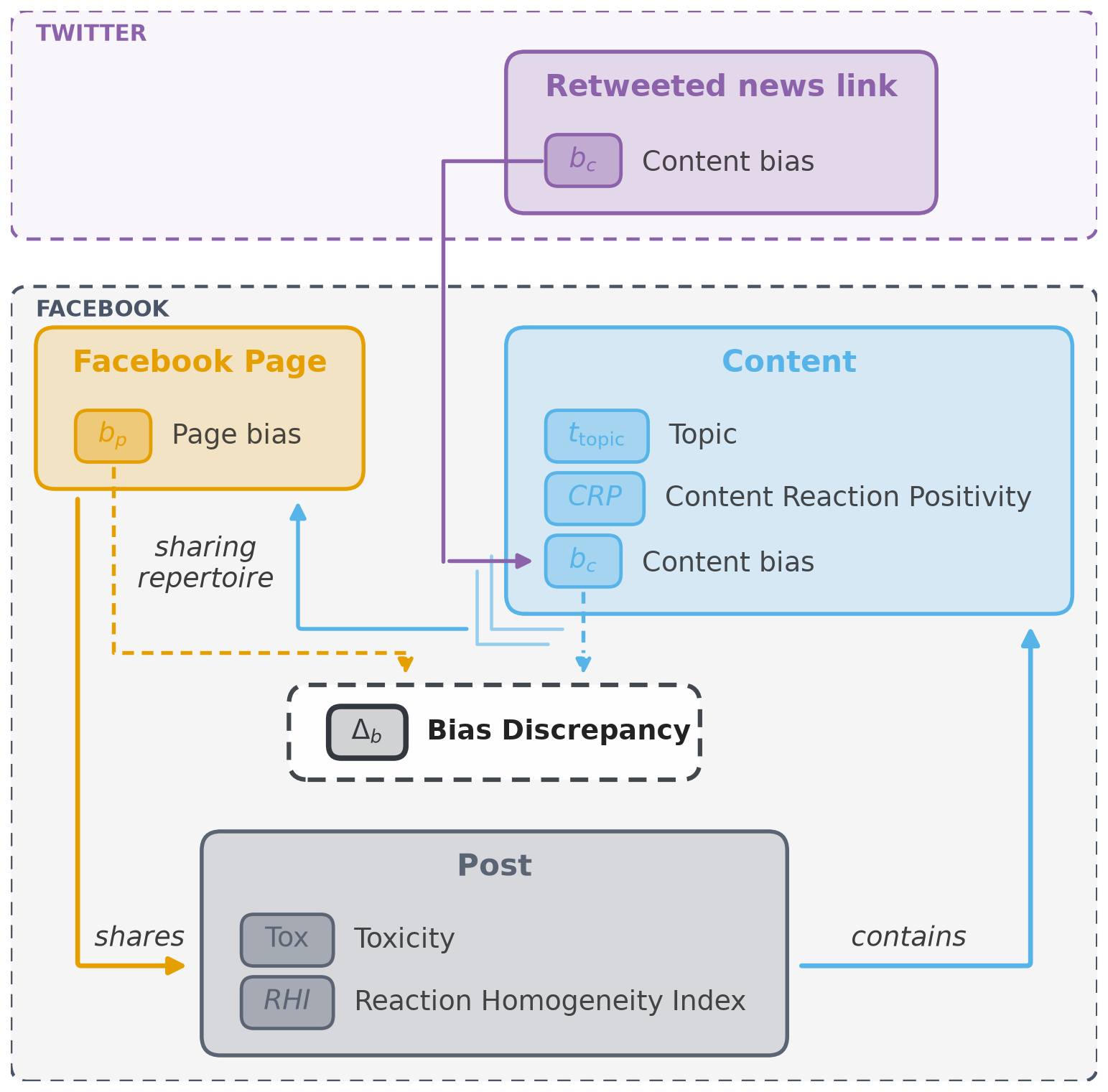}
\caption{Conceptual framework of the study. The dashed bands distinguish the Twitter and Facebook components of the analysis. On Twitter, each retweeted news link receives a content bias score ($b_c$) estimated from retweet behavior. The same score is attached to the corresponding content shared on Facebook and is the only measure transferred between platforms. On Facebook, each page is characterized by its page bias ($b_p$), which summarizes its ideological sharing repertoire and is estimated from the news links it shared. Individual posts connect pages to shared content and are characterized by toxicity and the Reaction Homogeneity Index (RHI). Content-level characteristics include $b_c$, topic, and Content Reaction Positivity (CRP), with CRP calculated from reactions to other posts sharing the same link. Bias Discrepancy is defined for each Facebook post as the absolute distance between content bias and page bias,
$\Delta_b = |b_c - b_p|$. Definitions and ranges of all measures are provided in the Methodology.}

\label{fig:framework}
\end{figure}

This study followed a six-stage analytical process: (1) data collection from
Facebook via the CrowdTangle API, (2) data preprocessing, (3) computation of
political bias metrics, (4) characterization of post text and shared news through toxicity scoring and news topic classification, (5) computation of audience reaction metrics, and (6) statistical modeling using a Generalized Linear Mixed Model (GLMM) with a Beta distribution. Fig~\ref{fig:framework} provides an overview of the analytical framework, illustrating the relationships between Facebook pages, the news links they share, and the posts that carry them, together with the Twitter-derived score used in the analysis. It shows how a page's estimated bias ($b_p$) and the bias of the shared news link ($b_c$) combine to produce Bias Discrepancy ($\Delta_b$), and it locates the other measures used in the analysis at the level of the content or individual post, including news topic, Content Reaction Positivity ($CRP$), toxicity, and the Reaction Homogeneity Index ($RHI$).

The framework combined information from two platforms with distinct roles. Twitter retweet data provided the ideological signal used to estimate the
political bias of each news link ($b_c$). These same news links were then identified in public Facebook posts, which constituted the observations
analyzed in this study. On Facebook, pages' sharing histories were used to estimate page bias ($b_p$), while individual posts provided the
accompanying text and audience reactions analyzed below.

\subsection*{Data collection}

\subsubsection*{Twitter data}

We used a set of $2,317$ unique political news links (URLs)
from the dataset associated with
\cite{kobellarz2022reaching}. These links originated from
Twitter data collected around the 2018 Brazilian presidential
election using election-related hashtags, covering the period
from October 1 to December 11, 2018.

Each link was associated with a precomputed political bias
score (originally denoted $RP(H)_{\text{content}}$ in
\cite{kobellarz2022reaching}, hereafter content bias, $b_c$), ranging
continuously from $-1.0$ (left) to $+1.0$ (right). Values
between $-0.3$ and $+0.3$ were considered center-leaning. These
scores were derived from the political polarity of Twitter
users who retweeted each news link, rather than from direct
ideological annotation of the news articles. The scores were
provided with the original dataset and were not independently
estimated in this study. A summary of their estimation and
validation procedures is provided in \nameref{app:TwitterBias},
with further details available in
\cite{kobellarz2022reaching}.

\subsubsection*{Facebook data}

Using the CrowdTangle API \cite{meta_crowdtangle_2024},
we collected public Facebook posts sharing the selected
news links. The collected information included post metadata,
textual content, and reaction counts, including Like, Love,
Angry, and Sad reactions.

Unlike the Twitter dataset, the Facebook collection did not
impose a separate publication-date window. Instead, we
searched for public Facebook posts sharing the selected
links, regardless of their publication dates.

\subsubsection*{Connecting Twitter and Facebook data}

The two datasets were connected through the URLs of the
shared news links. Each Facebook post was associated with
the precomputed content bias score ($b_c$)
of the corresponding link. Only this score crossed platforms, and no Twitter user-level information entered the Facebook analysis.

We used these scores as proxies for the ideological
orientation of the news links shared on Facebook. This
interpretation assumes that the ideological signal captured
through Twitter retweet behavior remains informative
across platforms, while recognizing that user populations
and interaction dynamics may differ between Twitter and
Facebook.

\subsection*{Data preprocessing}

Four filtering steps were applied in sequence to the 28,181 posts in the initial
collection, as summarized in Table~\ref{tab:data_cleaning}. First, duplicate
entries were removed. Second, we excluded observations lacking information
required for the analytical measures. This included posts with inaccessible
article text, missing content bias scores, no positive or negative
reactions among the four reaction types considered (Like, Love, Sad, and
Angry), or an undefined leave-one-out Content Reaction Positivity measure.
The latter occurred when the shared link had no other post with at least one
positive- or negative-valence reaction. Third, we excluded posts linking to three domains and two URL patterns (g1.globo.com, epoca.globo.com, jota.info, band.uol.com.br/eleicoes, and redetv.uol.com.br/aovivo) because automatic text extraction returned incomplete or non-article material, such as headlines, captions, registration notices, or navigation text. Finally,
posts with fewer than 50 total Facebook reactions were excluded. The minimum-reaction cutoff was determined through exploratory analysis of sample-size
stability and metric dispersion across increasing cutoff values. An empirical
assessment of this threshold selection is provided in~\nameref{app:cutoff}.

\begin{table}[htb]
\centering
\footnotesize
\caption{Data cleaning pipeline and post retention at each step.}
\label{tab:data_cleaning}
\begin{tabular}{llrr}
\toprule
\textbf{Step} & \textbf{Filtering procedure} &
\textbf{Remaining (N)} & \textbf{Loss (\%)} \\
\midrule
-- & Raw posts (initial collection) & 28,181 & -- \\
1 & Duplicates removed & 25,473 & 9.61\% \\
2 & Posts with incomplete analytical information removed &
17,973 & 29.44\% \\
3 & Posts from excluded domains removed &
11,424 & 36.44\% \\
4 & Posts with fewer than 50 reactions excluded &
4,584 & 59.87\% \\
\bottomrule
\end{tabular}
\end{table}

\subsection*{Analytical sample}

After all filtering steps, the final analytical sample comprised 4,584 Facebook
posts from 1,243 distinct pages sharing 761 distinct
news links. Their publication dates ranged from January 8,
2018 to November 3, 2022. Most posts were concentrated around the 2018 election: 4,341 (94.7\%) were published
in 2018, and 4,099 (89.4\%) fell within the Twitter collection window. Of these, 368 were published between
October 1 and the first-round vote, 3,031 between the
two rounds, and 700 between the second-round vote and
December 11.

Of the remaining posts, 222 (4.8\%) preceded the Twitter collection window, 20 (0.4\%) followed it within 2018, and 243 (5.3\%)
were published from 2019 to 2022. Although the Facebook data were not restricted to that window, the large majority of the analytical sample fell within it. Posts outside this period still shared links whose $b_c$ scores were estimated
from the 2018 Twitter data.

A page's sharing history comprises the posts it published in the deduplicated collection, before the filtering steps described above. $b_p$ was computed from the posts in this history whose shared link has a $b_c$ score. Sharing histories varied in size: the number of distinct links per page ranged from 1 to 288, with a median of 3 and a mean of 7.3. Of the 1,243 pages, 364 (29.3\%) had only one distinct link in their sharing history and contributed 371 posts (8.1\%) to the analytical sample.

The interval over which a page was represented in the analytical sample also varied with the number of posts observed. For 665 pages (53.5\%), the sample contained a single post and therefore defined no observation interval. Among the 578 pages with more than one post,
the median interval between the first and last observed posts was 21 days (interquartile range: 10.6--61.5 days),
and 63.7\% spanned 30 days or less. That is, for most pages observed more than once, the analytical sample covered weeks, not years. Publication dates, on the other hand, were not available for the full sharing history.

\subsection*{Political bias metrics}
\label{subsec:metrics}

To operationalize ideological discrepancy between pages and shared news links,
we defined two metrics: page bias ($b_p$) and Bias Discrepancy
($\Delta_b$).

We estimated each page's bias ($b_p$) by extending the polarization heuristic
introduced by \cite{kobellarz2022reaching} from hashtag-based to
content-sharing-based estimation. Specifically, $b_p$ summarizes the composition of a page's ideological sharing repertoire based on the
bias scores ($b_c$) of the news links it shared. Prior work evaluated the
original, hashtag-based heuristic in two political contexts, Brazil and
Canada, reporting substantial agreement among the independent annotators who
labeled the seed hashtags and retweet networks that are assortative by
estimated political orientation \cite{kobellarz2022reaching}. The link-based extension, however, is new to this study, and we assessed its face validity as described below.

Here, ``content'' refers to the shared news link, whose content bias score was derived from Twitter retweet networks. It does not include the text or commentary added by the page
alongside the link. The page bias score was computed as follows:

\begin{equation}
b_p = \frac{n_R w_R - n_L w_L}{n_L w_L + n_C w_C + n_R w_R},
\label{eqBc}
\end{equation}

where $n_L$, $n_C$, and $n_R$ denote the numbers of posts in the page's sharing history that share a news link classified as left-, center-, and right-leaning, respectively. The page-specific weights $w_R$, $w_L$, and $w_C$ are defined as
$w_R = \frac{n_R + n_C}{N}$,
$w_L = \frac{n_L + n_C}{N}$, and
$w_C = \frac{n_R + n_L}{N}$, 
where $N = n_R + n_L + n_C$ denotes the total number of such posts.

When all of them were classified as center-leaning
($n_L = n_R = 0$), the denominator of Eq.~\ref{eqBc} is zero. We therefore defined $b_p = 0$ for this degenerate case. Values of $b_p$ approaching $+1.0$ indicate an ideological sharing repertoire dominated by right-leaning content, whereas values approaching
$-1.0$ indicate a repertoire dominated by left-leaning content.

The informativeness of $b_p$ as a measure of a page's ideological sharing
repertoire depends on the number of distinct links observed for that page. In
the limiting case of a page represented by only one distinct link, $b_p$ is
determined entirely by the ideological category of that link and can therefore
take only $-1$, $0$, or $+1$ under our weighting scheme. In such cases, $b_p$
should not be interpreted as a stable estimate of the page's broader
ideological orientation, but rather as a characterization of its observed
sharing behavior within the available data. We assessed this limitation through a sensitivity analysis that excluded pages
represented by only one distinct link, reported in
\nameref{app:r27_checks}.

To assess the face validity of $b_p$, we examined all 117 pages in the
analytical sample with at least 15 distinct shared links. We assigned Left,
Center, or Right labels from the identity of named political actors or
movements, explicit ideological self-descriptions, and the editorial identity
of recognizable media outlets, without consulting $b_p$. This yielded labels
for 66 pages, while the other 51 could not be classified from their identities.
We then categorized $b_p$ using cutoffs at $-0.3$ and $+0.3$. The
identity-based and $b_p$-based categories matched for 62 of the 66 classified
pages (93.9\%), and the continuous $b_p$ scores tracked the ordered
identity-based labels (Spearman's $\rho=0.897$). Agreement was lower for media
outlets (15/18), whose labels involved editorial judgment. The coding rules and
category-level results are reported in
\nameref{app:publisher_validation}.

With $b_p$ operationalized, we computed the Bias Discrepancy ($\Delta_b$), defined in Eq.~\ref{eq:deltabias}, to quantify how far a particular shared link lies from a page's ideological
sharing repertoire.
%
%
%
Values near $0$ indicate that the link is similar in ideological orientation
to that repertoire, while larger values indicate a greater departure from it. Because $b_p$ was estimated from sharing history and not from an independent assessment of the page's
ideology, $\Delta_b$ measures how far a share falls from that
repertoire, not its incongruence with an established page position. It cannot
distinguish, on its own, whether such a share reflects surprise, criticism,
attention seeking, or another motive. We computed one $\Delta_b$ value for each Facebook post and examined its association with audience reactions and toxicity.

\subsection*{Toxicity and topic characterization}
\label{subsec:content}

We also characterized each Facebook post with two measures taken from different texts. First, we analyzed the linguistic tone of the text written by the page alongside the
shared link using toxicity scoring. Second, we scored the linked news article on ten topics using zero-shot classification.

Toxicity ($\text{Tox}$) captures the presence of hostile or inflammatory
language in the text written by the page alongside the shared news link. To measure it, we employed Google's Perspective API \cite{perspective_api_2024}, which returns a
continuous score $\text{Tox} \in [0,1]$ indicating the likelihood that a text
may be perceived as toxic (e.g., insults, threats, or hate speech). We
processed the Facebook post text directly in its original language
(Portuguese), rather than translating it into English. This decision followed findings by \cite{KobellarzWebMedia2022}, who showed that automatic translation
from Brazilian Portuguese to English can artificially reduce toxicity scores,
particularly for highly toxic comments.

To characterize the thematic content of the shared news articles, we employed
the multilingual transformer model mDeBERTa-v3-base for zero-shot topic
classification. The procedure used to define the topic categories and select
the final classification model is described in~\nameref{app:topicSel}. The resulting topic scores, one per topic and article, allowed us to account for
differences in the subject of the news when modeling reaction homogeneity.

\subsection*{Audience reaction metrics}
\label{subsec:engagement}

We next defined two metrics derived from Facebook reaction counts. They
describe different aspects of audience response: Content Reaction
Positivity ($CRP$) characterizes how a shared news link is received across
other posts carrying the same link, whereas the Reaction Homogeneity Index
($RHI$) characterizes the balance of positive and negative reactions on the
focal post.

Reaction counts on Facebook provide observable signals of audience responses
to political content \cite{de2021sadness}. We classified Like and Love as positive-valence reactions and Sad and Angry as negative-valence reactions. Haha and Wow were excluded because their valence is more context dependent and
cannot be mapped consistently onto a positive--negative axis. The four reaction types retained in the analysis accounted for 82.09\% of all reactions in the
analytical sample. To characterize how the shared news
link was received in other posts, we defined Content Reaction Positivity ($CRP$) from reactions to all other posts sharing link $c$, excluding the
focal post $i$. Let $\mathcal{P}_c$ denote the set of posts sharing $c$. Then:

\begin{equation*}
CRP_i =
\dfrac{\displaystyle\sum_{\substack{j \in \mathcal{P}_c \\ j \neq i}} n_j^+}
{\displaystyle\sum_{\substack{j \in \mathcal{P}_c \\ j \neq i}}
\left(n_j^+ + n_j^-\right)},
\end{equation*}

where $n_j^+$ and $n_j^-$ denote the counts of positive reactions (Like, Love)
and negative reactions (Angry, Sad) for post $j$. The score ranges from $0$ to
$1$, with higher values indicating predominantly positive reactions to the
shared link in other posts. $CRP_i$ describes reactions to the same link
elsewhere in the dataset, not reactions to the focal post.

For a focal post to have a defined leave-one-out Content Reaction Positivity
score, the same news link must be represented by at least one other post with
at least one positive- or negative-valence reaction. Observations for which no
such comparison post was available were excluded during preprocessing.

To characterize the balance of positive and negative reactions on the focal
post, we defined the Reaction Homogeneity Index ($RHI$):

\begin{equation*}
RHI_i = \frac{|n_i^+ - n_i^-|}{n_i^+ + n_i^-}.
\end{equation*}

Values close to $1$ indicate that reactions are predominantly of one sign,
whether positive or negative, whereas values near $0$ indicate that positive
and negative reactions occur in similar numbers. The index characterizes the
distribution of reaction types and does not measure agreement among users,
their political positions, or political polarization.

Both measures were assigned to each focal post but were computed from different sets of reactions. $CRP_i$ summarizes reactions to other posts sharing the
same news link ($j \neq i$), whereas $RHI_i$ uses only reactions to the focal
post $i$. A post with predominantly negative reactions can therefore have a
high $RHI_i$, as can a post with predominantly positive reactions, and a
roughly even mix produces a low $RHI_i$. If
$f_i^- = n_i^-/(n_i^+ + n_i^-)$ denotes the focal post's negative-reaction
share, then $RHI_i = |1 - 2f_i^-|$. In this sample, 4,347 posts (94.8\%)
received at least as many positive as negative reactions ($f_i^- \leq 0.5$),
so $RHI_i = 1 - 2f_i^-$ for these observations. In other words, for the large majority of posts, lower $RHI$ is mathematically a larger negative-reaction share. The
negative-share analyses in \nameref{app:r27_checks} are therefore another
view of the same reaction-composition pattern, not independent
corroborating evidence. By contrast, no such identity holds between $CRP_i$ and $RHI_i$,
which are computed from different posts. Their observed correlation ($r \approx 0.75$) is therefore an empirical association and not evidence that
reaction positivity causes reaction homogeneity.

\subsection*{Statistical modeling}
\label{subsec:glmm}

\subsubsection*{Main model}

To address RQ2, we estimated the association between Bias Discrepancy and
reaction homogeneity while accounting for other page, content, and post
characteristics. Our main model was a Generalized Linear Mixed Model
(GLMM) with a Beta distribution and logit link.

The fixed effects included signed content and page bias ($b_c$ and $b_p$),
their absolute values ($|b_c|$ and $|b_p|$), Bias Discrepancy
($\Delta_b$), toxicity, Content Reaction Positivity ($CRP$), ten topic
scores, and controls for total reactions and page links. Signed bias
captures left--right direction, whereas absolute bias captures distance
from the ideological center. Including these terms alongside
$\Delta_b$ allowed us to estimate the discrepancy association while
accounting for the direction and extremity of page and content bias.
This representation also provided the best fit among the bias representations considered and was retained in the
main model.

The topic scores were included as separate continuous predictors because
a news article may be associated with multiple topics. The activity controls were log-transformed: total reactions included all reaction types recorded for the focal post, and page links counted distinct links in the
page's sharing history. By contrast, the valence-based reaction measures used only Like, Love, Sad, and Angry. Finally, including $CRP$ accounted for reactions to the same link in other posts. Its coefficient
is read as an empirical association, not as an independent behavioral
effect.

The data had a cross-classified structure: pages could contribute multiple posts, and the same news link could be shared by multiple pages. We therefore included random intercepts for pages and content (shared news links).
Random slopes for toxicity at both levels allowed its association with
reaction homogeneity to vary across pages and links.

Because the Beta distribution is defined on $(0,1)$, we transformed the observed $RHI$ values, including those at $0$ and $1$, into the open
interval using the adjustment proposed by Smithson and
Verkuilen~\cite{smithson2006better}. The transformed outcome is denoted
$RHI'$, with details provided in \nameref{app:beta}.

The main model was estimated using \texttt{glmmTMB} in R:

\begin{equation*}
\begin{aligned}
RHI' \sim\;& b_c + |b_c| + b_p + |b_p| + \Delta_b
          + \text{Tox} + CRP \\
&+ \log(\text{total reactions})
 + \log(\text{page links}) \\
&+ t_{\text{Economy}} + t_{\text{Education}}
 + t_{\text{Health}} + t_{\text{Security}} \\
&+ t_{\text{Culture}} + t_{\text{Religion}}
 + t_{\text{Disinformation}} \\
&+ t_{\text{Election}} + t_{\text{Politics}}
 + t_{\text{Corruption}} \\
&+ (\text{Tox} \mid \text{Page})
 + (\text{Tox} \mid \text{Content}).
\end{aligned}
\end{equation*}

Here, each $t_{\text{topic}} \in [0,1]$ is the zero-shot classifier score
assigned to that topic for the shared news link.

\subsubsection*{Estimation and diagnostics}

All continuous predictors were standardized using Z-score normalization within each fitted sample, and the boundary adjustment of $RHI$ was recomputed using that sample's size. Because the continuous predictors were standardized, we compared their associations with $RHI$ by the
absolute size of their coefficients. Coefficient confidence intervals were two-sided 95\% Wald intervals. Marginal and conditional $R^2$ were calculated using the Nakagawa method implemented in
\texttt{performance::r2\_nakagawa()}, with the lognormal approximation.

Because $\Delta_b$ is derived from $b_p$ and $b_c$, we assessed potential multicollinearity using \texttt{performance::check\_collinearity()}.
All VIFs were below 5 in the main model, and the maximum across the sensitivity models was 5.10.

\subsubsection*{Sensitivity analyses}

We compared three bias representations: signed content and page bias
alone, signed bias together with absolute extremity, and Left/Center/Right
categories. These comparisons used the same posts and remaining covariates but omitted the two activity controls.

We also refit the main model after omitting $CRP$, restricting pages to
at least two or five distinct bias-scored links, restricting posts to
the Twitter collection window, recalculating page bias without any
shares of the focal link by the page, and replacing the minimum-reaction
cutoff of 50 with 25 or 100.

Finally, we analyzed the focal post's negative-reaction share,
$f_i^- = n_i^-/(n_i^+ + n_i^-)$, using a mixed beta model and a
beta-binomial model of the underlying reaction counts. Both retained the main model's other terms while omitting $CRP$. Full model-comparison and sensitivity details are reported in \nameref{app:r27_checks}.

\section*{Results}
\label{cap:results}

We first describe the sample and the distributions of the analytical
variables. Because Bias Discrepancy combines page bias and content bias, we
then examine how each of the two relates to audience reactions and toxicity,
before addressing RQ1 through the analysis of Bias Discrepancy and RQ2 through
the main model.

\subsection*{Sample composition and variable distributions}

Table~\ref{tab:political_orientation} shows that both shared news links and pages were skewed toward the left. Most posts shared left-leaning links (52.79\%), followed by center-leaning
(31.48\%) and right-leaning links (15.73\%). Pages were also predominantly left-leaning, although the ordering differed: left-leaning pages accounted for 54.55\% of the total, right-leaning pages represented 25.91\%, and center-leaning pages were the least frequent group (19.55\%).

\begin{table}[htb]
\centering
\footnotesize
\caption{Distribution of content bias categories across Facebook posts and page bias
categories across distinct pages.}
\label{tab:political_orientation}
\begin{tabular}{lrr}
\toprule
\textbf{Category} & \textbf{N} & \textbf{Percentage} \\
\midrule
\multicolumn{3}{c}{\textit{Content bias (N = 4,584 posts)}} \\
\midrule
Left ($b_c \leq -0.3$)       & 2,420 & 52.79\% \\
Center ($-0.3 < b_c < 0.3$)  & 1,443 & 31.48\% \\
Right ($b_c \geq 0.3$)       & 721   & 15.73\% \\
\midrule
\multicolumn{3}{c}{\textit{Page bias (N = 1,243 pages)}} \\
\midrule
Left ($b_p \leq -0.3$)       & 678 & 54.55\% \\
Center ($-0.3 < b_p < 0.3$)  & 243 & 19.55\% \\
Right ($b_p \geq 0.3$)       & 322 & 25.91\% \\
\bottomrule
\end{tabular}
\end{table}

\begin{figure}[htb]
    \centering
    \includegraphics[width=.8\linewidth]{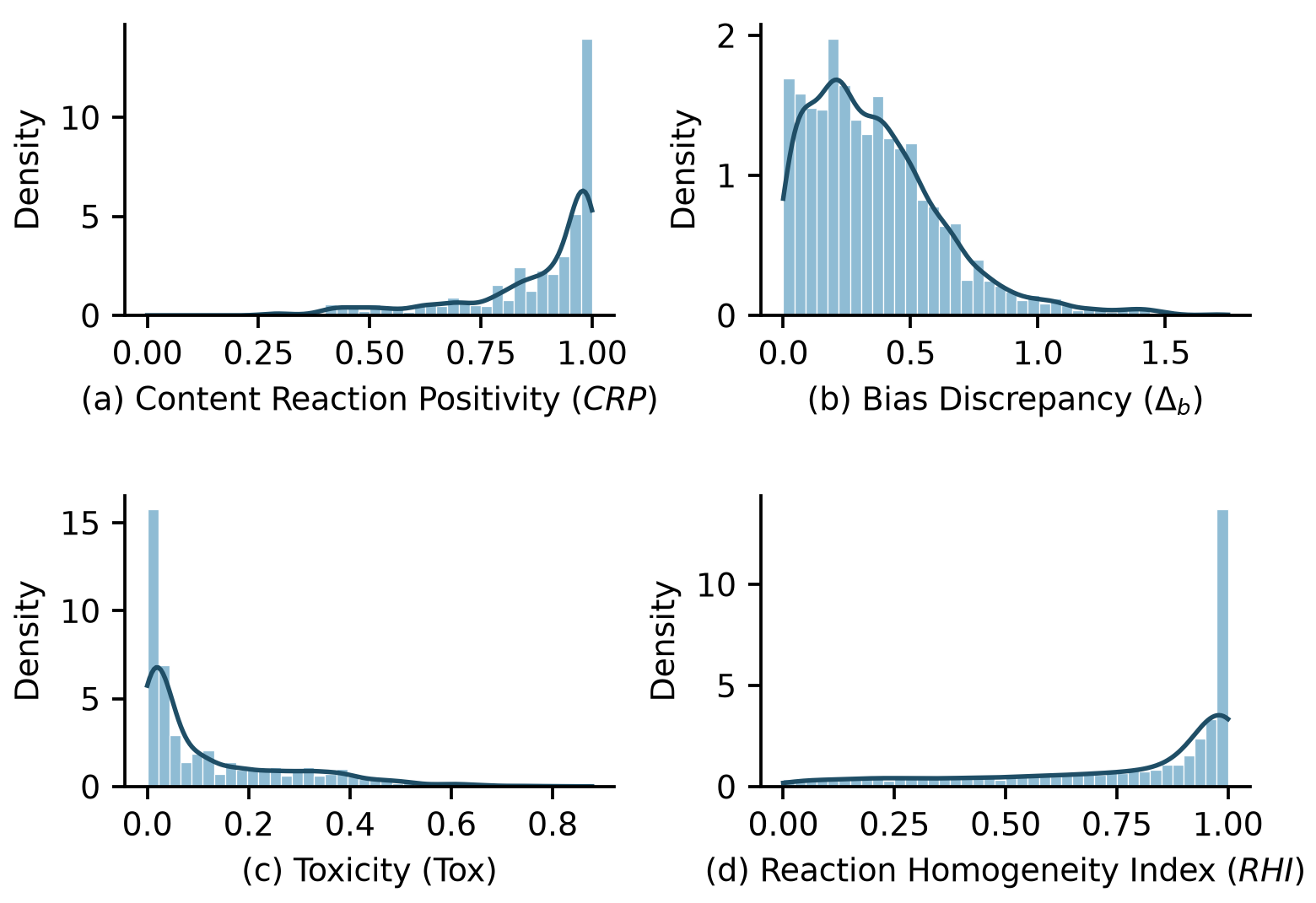}
    \caption{Distributions of the main analytical variables. Curves represent
    kernel density estimates.}
    \label{fig:metricsHistogram}
\end{figure}

Fig~\ref{fig:metricsHistogram} shows the distributions of the main analytical
variables. Content Reaction Positivity ($CRP$) was concentrated near $1.0$, indicating that reactions to shared links in other posts were predominantly positive (Like, Love). Bias Discrepancy ($\Delta_b$) was concentrated at lower values, indicating that most posts shared content
relatively close to the page's ideological sharing repertoire.
Toxicity ($\text{Tox}$) had a right-skewed distribution, with most posts
having low toxicity scores and a smaller number extending toward higher
values. Finally, Reaction Homogeneity Index ($RHI$) values were concentrated near $1.0$, indicating that reactions on most posts were predominantly of one sign rather than evenly divided between positive and negative reactions.

\begin{figure}[ht]
  \centering
  \includegraphics[width=.8\linewidth]{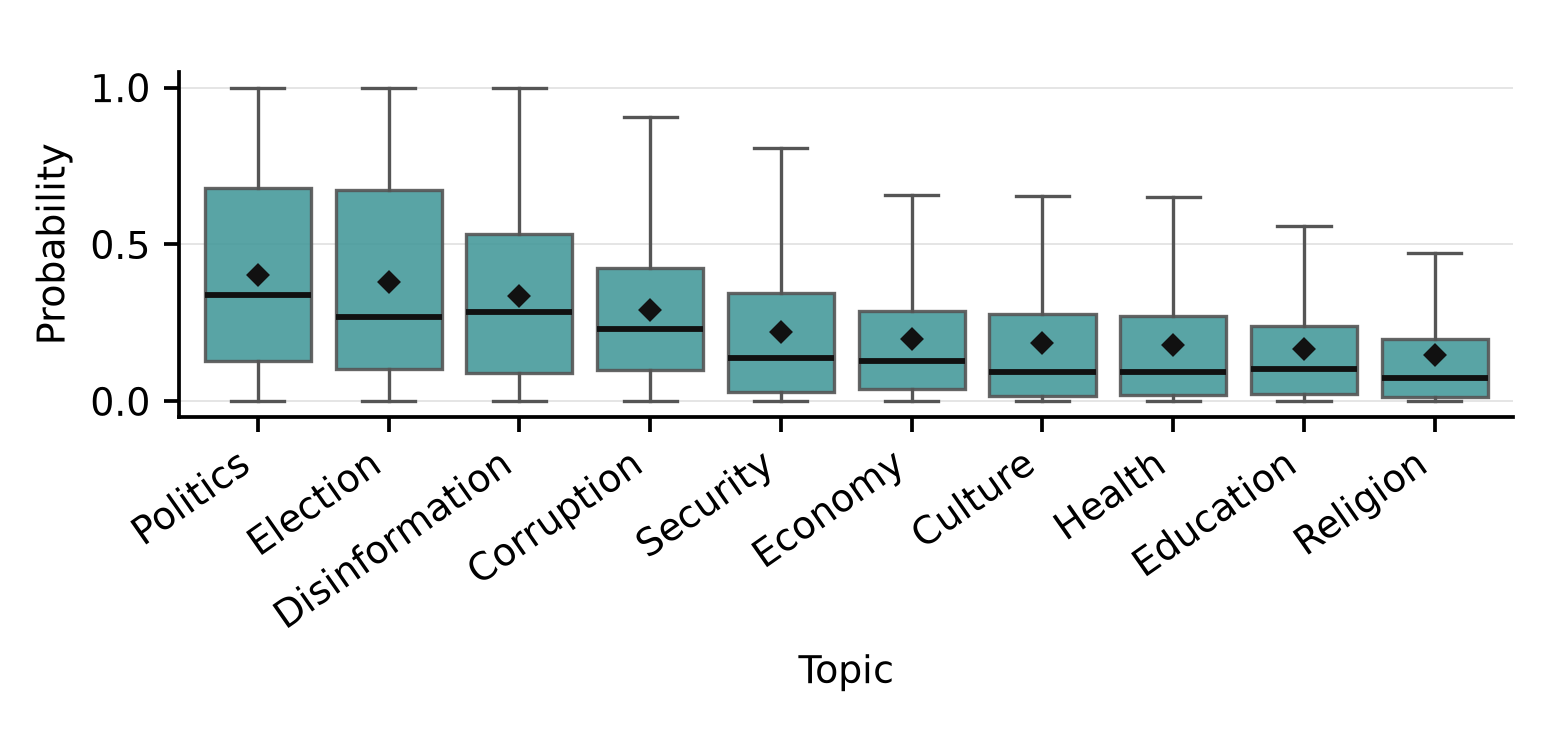}
  \caption{Distribution of zero-shot classification scores across topics.
  Boxplots represent the score assigned to each topic across shared news
  links.}
  \label{fig:topic_probabilities}
\end{figure}

Fig~\ref{fig:topic_probabilities} shows the distribution of topic scores
across the shared news links. Politics and elections
($t_{\text{Politics}}$, $t_{\text{Election}}$) had the highest median scores and wide variability, indicating that these topics were common across the corpus. In comparison, education and religion ($t_{\text{Education}}$, $t_{\text{Religion}}$) had lower median scores, although high-score observations indicated that they were prominent in some news articles.

\subsection*{Association between political bias and audience reactions}

\begin{figure}[ht]
    \centering
    \includegraphics[width=.8\linewidth]{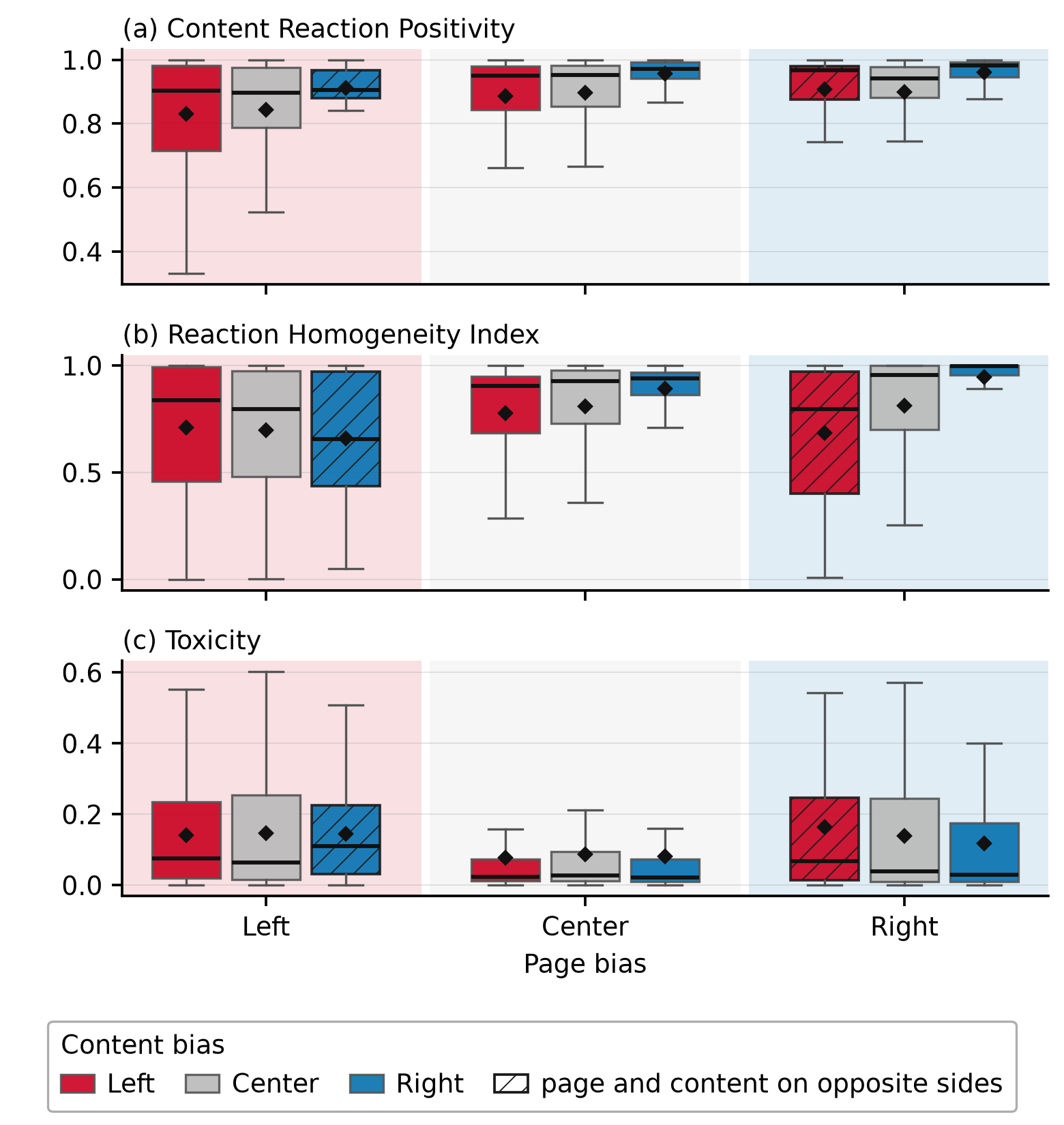}
    \caption{Distribution of Content Reaction Positivity, Reaction Homogeneity
    Index, and Toxicity across page and content bias categories. Panels (a) to
    (c) show Content Reaction Positivity, the Reaction Homogeneity Index, and
    Toxicity, respectively. The background band of each group indicates page
    bias and the box fill indicates content bias. Boxes whose fill contrasts
    with the background band represent posts in which page and content fall on
    opposite ideological sides and are shown with hatching.}
    \label{fig:metrics_comparison}
\end{figure}

Fig~\ref{fig:metrics_comparison} illustrates how page bias ($b_p$) and content bias ($b_c$) were associated with Content Reaction Positivity, the
Reaction Homogeneity Index, and the toxicity of the accompanying post text.
Statistical significance was assessed using Welch's ANOVA and Games--Howell
post-hoc tests to account for unequal variances. 

\textit{Content Reaction Positivity
(Fig~\ref{fig:metrics_comparison}, panel (a)):}
Content Reaction Positivity varied systematically across ideological groups. Links shared by left-leaning pages had lower average Content Reaction
Positivity than those shared by center- and right-leaning pages
($p < 0.001$). Across content categories, right-leaning links had higher average
Content Reaction Positivity than left- and center-leaning links
($p < 0.05$).

\textit{Reaction Homogeneity Index
(Fig~\ref{fig:metrics_comparison}, panel (b)):}
Reaction homogeneity varied across combinations of page and content bias. For
center-leaning content, posts by left-leaning pages had lower $RHI$ than posts
by center- and right-leaning pages (both $p < 0.001$).
For right-leaning content, $RHI$ increased from left-
to center- to right-leaning pages, with all pairwise
differences significant ($p < 0.001$).
Within left-leaning pages, however, $RHI$ did not differ significantly across content-bias categories
($p = 0.286$).

The lowest observed mean $RHI$ occurred when left-leaning pages shared right-leaning content, although, as noted above, differences across content-bias categories within left-leaning pages were not statistically significant.
Posts by right-leaning pages sharing left-leaning content
also had lower mean $RHI$ than those sharing right-leaning content ($p < 0.001$).
These comparisons describe the distribution of reaction
types and do not establish whether individual users
encountered opposing views or disagreed about the
political meaning of the content.

\textit{Toxicity (Fig~\ref{fig:metrics_comparison}, panel (c)):}
Toxicity patterns differed from those observed for Content Reaction Positivity
and the Reaction Homogeneity Index. Center-leaning pages had lower toxicity than both left- and right-leaning pages across content categories
(all $p < 0.01$). With respect to content bias, we found no consistent evidence that posts sharing ideologically opposing content contained more toxic
text than posts sharing aligned content, despite variation in median toxicity
across groups.

Page and content bias were therefore associated with differences in Content
Reaction Positivity and reaction homogeneity.
Among these differences, posts by right-leaning pages generally had more homogeneous reactions in this sample. This pattern is compatible with prior work reporting stronger cohesion
within right-leaning communities in the Brazilian context
\cite{Kobellarz2019,kobellarz2022reaching}, although reaction homogeneity in
our data does not itself measure audience cohesion or political agreement.

\subsection*{Association between Bias Discrepancy, audience reactions, and toxicity}

The comparisons above used bias categories. Bias Discrepancy instead
measures, on a continuous scale, how far a shared link lies from the page's
ideological sharing repertoire. To examine how it related to audience
reactions and toxicity (RQ1), we compared Content Reaction Positivity
($CRP$), the Reaction Homogeneity Index ($RHI$), and Toxicity
($\text{Tox}$) across quartiles of Bias Discrepancy ($\Delta_b$), from
``Very Low'' (strong alignment with that repertoire) to ``Very High''
(strong departure from it). Statistical significance was assessed using
Welch's ANOVA and Games--Howell post-hoc tests.

The quartile cut points for $\Delta_b$ were $0.161$, $0.313$, and $0.498$.
The four groups contained 1,148, 1,148, 1,142, and 1,146 posts, respectively,
and their similar sizes follow from the quartile definition. Each group included hundreds of distinct pages and news links, indicating that the observed discrepancy patterns were not confined to a small number of pages or news links. The Very High discrepancy group, for instance, included 595
distinct pages, with no single page accounting for more than 3.9\% of posts
in that group. Page concentration by group is reported in
\nameref{app:discrepancy_distribution}.

\begin{figure}[ht]
    \centering
    \includegraphics[width=.8\linewidth]{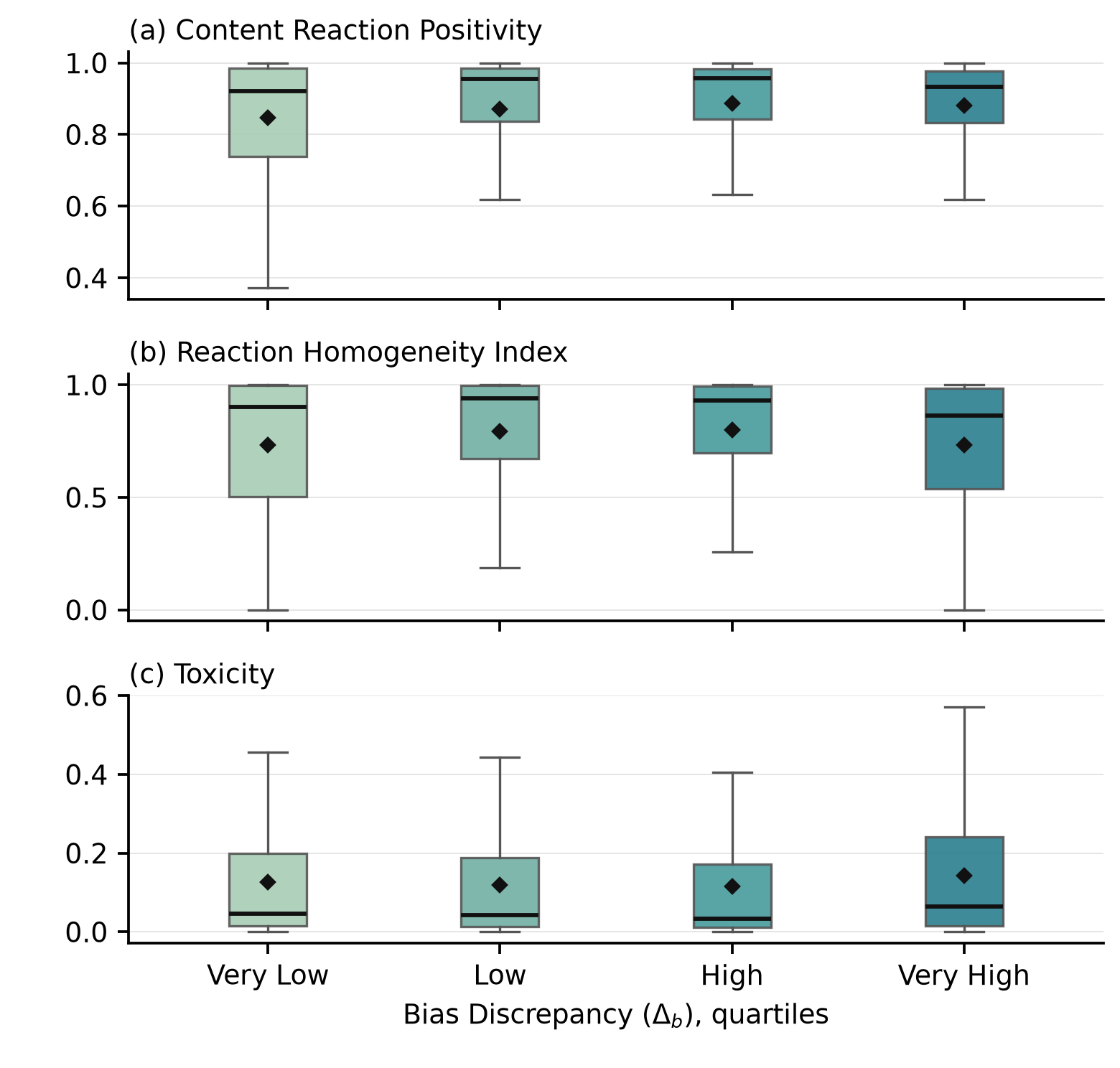}
    \caption{Distribution of Content Reaction Positivity, Reaction Homogeneity
    Index, and Toxicity across quartiles of Bias Discrepancy ($\Delta_b$), from
    ``Very Low'' to ``Very High.'' Panels (a) to (c) show Content Reaction
    Positivity, the Reaction Homogeneity Index, and Toxicity, respectively.}
    \label{fig:delta_analysis}
\end{figure}

\textit{Content Reaction Positivity
(Fig~\ref{fig:delta_analysis}, panel (a)):}
Content Reaction Positivity was significantly lower in the ``Very Low'' Bias
Discrepancy category than in all other levels of Bias Discrepancy
($p < 0.001$). Put differently, links closely aligned with a page's ideological sharing repertoire tended to receive a lower proportion of positive reactions in other
posts sharing the same link. The data do not show what produced this
unexpected pattern, and we interpret it with caution.

\textit{Reaction Homogeneity Index
(Fig~\ref{fig:delta_analysis}, panel (b)):}
Reaction homogeneity was significantly lower in the ``Very Low'' and ``Very
High'' Bias Discrepancy categories than in the ``Low'' and ``High''
categories. No statistically significant difference was observed between the
``Very Low'' and ``Very High'' groups. The relationship between Bias
Discrepancy and reaction homogeneity was therefore non-monotonic, with lower homogeneity
at both ends of the discrepancy distribution.

\textit{Toxicity (Fig~\ref{fig:delta_analysis}, panel (c)):}
Toxicity also differed across Bias Discrepancy levels ($p < 0.001$). No statistically significant differences were observed among the lower and
moderate discrepancy groups. The ``Very High'' category, however, had significantly higher toxicity than all other groups. This does not conflict with the
category comparison above, which found no consistent toxicity difference
between posts sharing opposite-side and aligned content. Every post pairing
a page and a link on opposite sides fell in the Very High group, but these were only 176 of its 1,146 posts. Most of the group paired a center-leaning
page with a left- or right-leaning link, or the reverse (733 posts), and the rest were large departures within the same side (237 posts).

The three outcomes thus related to Bias Discrepancy in different ways. For
RQ1, the greatest departures from a page's ideological sharing
repertoire were associated with less homogeneous reactions and more toxic accompanying text, whereas the lowest Content Reaction Positivity occurred at
the opposite end of the discrepancy distribution, where reaction homogeneity was also lower than in the middle quartiles.

\subsection*{Modeling factors associated with reaction homogeneity}
\label{secReModel}

The quartile comparisons in the preceding section describe the unadjusted relationship between Bias Discrepancy and reaction homogeneity, with lower $RHI$ at both ends of the observed discrepancy distribution. The mixed model below answers a different question. It estimates the association between continuous $\Delta_b$ and $RHI$ conditional on the remaining covariates and on clustering by Facebook page and shared news link. Because $\Delta_b$ enters the model as a linear term on the logit scale, its coefficient summarizes an adjusted association across the observed discrepancy range. It does not imply that the unadjusted relationship is monotonic.

Table~\ref{tab:beta_results} presents the main model predicting the
Reaction Homogeneity Index, which separates signed bias from absolute
extremity and controls for total reactions and page links. Its marginal and conditional $R^2$ were $0.762$ and $0.977$, respectively.

\begin{table}[htb!]
\centering
\scriptsize
\caption{Main model: mixed beta regression predicting the Reaction Homogeneity Index.
Continuous predictors are standardized. The model includes page and content
random intercepts and random slopes for toxicity.}
\label{tab:beta_results}
\begin{tabular}{lccc}
\toprule
Predictor & $\beta$ & 95\% CI & $p$ \\
\midrule
Intercept & 1.536 & 1.464 -- 1.607 & $<0.001$ \\
Content bias ($b_c$) & 0.158 & 0.081 -- 0.236 & $<0.001$ \\
Content extremity ($|b_c|$) & 0.072 & 0.002 -- 0.143 & 0.044 \\
Page bias ($b_p$) & 0.099 & 0.043 -- 0.155 & $<0.001$ \\
Page extremity ($|b_p|$) & 0.114 & 0.059 -- 0.168 & $<0.001$ \\
Bias Discrepancy ($\Delta_b$) & -0.173 & -0.211 -- -0.135 & $<0.001$ \\
Toxicity ($\text{Tox}$) & -0.149 & -0.196 -- -0.102 & $<0.001$ \\
Content Reaction Positivity ($CRP$) & 0.951 & 0.888 -- 1.013 & $<0.001$ \\
$\log(\text{total reactions})$ & -0.115 & -0.159 -- -0.071 & $<0.001$ \\
$\log(\text{page links})$ & -0.071 & -0.126 -- -0.017 & 0.010 \\
$t_{\text{Economy}}$ & 0.034 & -0.045 -- 0.114 & 0.397 \\
$t_{\text{Education}}$ & 0.113 & 0.032 -- 0.193 & 0.006 \\
$t_{\text{Health}}$ & -0.096 & -0.190 -- -0.003 & 0.044 \\
$t_{\text{Security}}$ & 0.079 & -0.023 -- 0.181 & 0.128 \\
$t_{\text{Culture}}$ & -0.093 & -0.203 -- 0.017 & 0.098 \\
$t_{\text{Religion}}$ & -0.018 & -0.099 -- 0.063 & 0.658 \\
$t_{\text{Disinformation}}$ & -0.101 & -0.195 -- -0.006 & 0.036 \\
$t_{\text{Election}}$ & -0.011 & -0.110 -- 0.088 & 0.831 \\
$t_{\text{Politics}}$ & 0.085 & -0.026 -- 0.197 & 0.134 \\
$t_{\text{Corruption}}$ & 0.017 & -0.087 -- 0.122 & 0.743 \\
\midrule
Posts / pages / links & \multicolumn{3}{c}{4,584 / 1,243 / 761} \\
Marginal / conditional $R^2$ & \multicolumn{3}{c}{0.762 / 0.977} \\
AIC & \multicolumn{3}{c}{-15046.2} \\
\bottomrule
\end{tabular}
\end{table}

Greater Bias Discrepancy was associated with lower reaction homogeneity ($\beta=-0.173$, $p<0.001$), as was toxicity
($\beta=-0.149$, $p<0.001$). By contrast, Content Reaction Positivity ($CRP$), which summarizes reactions to the same link in other posts, was positively associated with $RHI$ ($\beta=0.951$). Both activity controls had negative coefficients: posts with more reactions and posts from pages sharing more distinct links tended to have lower $RHI$, conditional on the other model terms.

For RQ2, Content Reaction Positivity therefore had the largest standardized
coefficient. Bias Discrepancy had the next largest, followed by content bias
($\beta=0.158$) and toxicity. The confidence intervals of these three
overlapped, so their relative order was not well determined.

Across the main model and seven sensitivity analyses, the
standardized Bias Discrepancy coefficient remained negative, ranging
from $-0.219$ to $-0.167$, with all 95\% confidence intervals excluding
zero. The association persisted when $CRP$ was omitted
($\beta=-0.178$, 95\% CI $[-0.219,-0.138]$) and when page bias was recalculated
without the focal shared link ($\beta=-0.192$, 95\% CI $[-0.234,-0.150]$).
It also persisted under the restrictions on bias-scored links,
restriction to the Twitter collection window, and alternative minimum-reaction cutoffs. Toxicity likewise retained a negative coefficient in all 
specifications, with all 95\% confidence intervals excluding zero.
Full results are reported in \nameref{app:r27_checks}.

To provide concrete illustrations of the relationship between Bias
Discrepancy and reaction homogeneity, we examined high-discrepancy observations
from pages represented by at least 15 distinct links and selected examples in
both ideological directions. For \textit{A Gazeta Esquerdista Contra o
Coiso}, whose ideological sharing repertoire was strongly left-leaning
($b_p=-0.833$), sharing a right-leaning link ($b_c=+0.615$) produced a large
Bias Discrepancy ($\Delta_b=1.449$). The post had an $RHI$ of $0.135$, below
the page's mean $RHI$ of $0.739$.

Conversely, \textit{Grupo do Movimento Brasil à Direita} had a
right-leaning ideological sharing repertoire ($b_p=+0.580$) but shared a
left-leaning link ($b_c=-0.749$), resulting in a high Bias Discrepancy
($\Delta_b=1.329$). The page shared this same link in two separate posts,
which drew different reactions. The post with
higher toxicity ($\text{Tox}=0.47$) had a relatively
low $RHI$ of $0.375$, compared with the page-level mean of $0.842$
(median $=1.000$). The other post, with a toxicity of $0.01$, had an
$RHI$ of $0.944$.

These two examples show that high Bias Discrepancy can coincide with
relatively heterogeneous reactions in either ideological direction. The two
posts from \textit{Grupo do Movimento Brasil à Direita} also allow a
within-page comparison. With the page and the link held fixed, the post with
the more toxic accompanying text drew less homogeneous reactions.

A counter-example illustrates that this pattern does not apply to every
high-discrepancy post. \textit{Brasil Conservador} ($b_p=+0.681$) shared a
strongly left-leaning link ($b_c=-0.749$), producing a similarly large Bias
Discrepancy ($\Delta_b=1.430$), yet the post had highly homogeneous reactions
($RHI=0.960$), close to the page's mean $RHI$ of $0.951$. The negative
association between Bias Discrepancy and $RHI$ should therefore be interpreted
as an aggregate tendency rather than a post-level rule.

\begin{figure}[htb!]
    \centering
    \includegraphics[width=.9\linewidth]{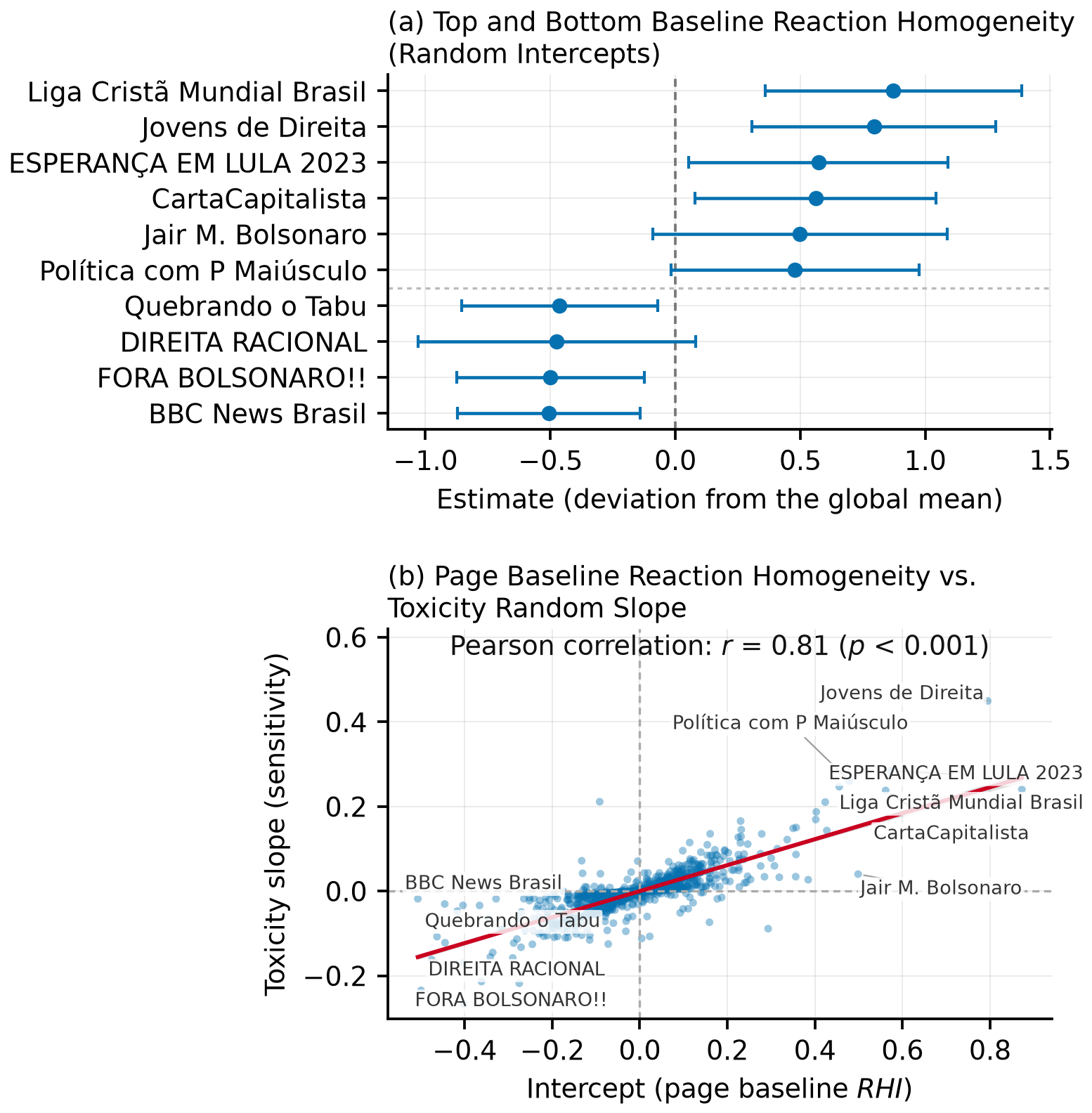}
    \caption{Page-level heterogeneity from the main model. Panel (a)
    shows pages with the most extreme random-intercept deviations in baseline
    Reaction Homogeneity Index, with 95\% intervals. Panel (b) relates
    page-level random intercepts to random slopes for toxicity. The two are
    positively correlated ($r=0.81$).}
    \label{fig:account_effects}
\end{figure}

Beyond the fixed effects, the random effects describe how pages differed from
one another.
Fig~\ref{fig:account_effects}, panel (a), shows the pages with the
largest deviations in baseline reaction homogeneity (random intercepts).
Pages with strongly negative deviations included both news-oriented pages
(e.g., ``BBC News Brasil'') and politically oriented pages, indicating that lower baseline reaction homogeneity was not exclusive to a single type of page.
The presence of a major news outlet among these cases is consistent with prior
work describing some news outlets as \textit{bubble reachers} (or brokers),
capable of distributing content across different ideological clusters
\cite{kobellarz2022reaching}. Such broader reach could be associated with more
heterogeneous audience reaction patterns, although the random effects
estimated here do not directly measure audience composition or brokerage.

In contrast, several pages with strongly positive deviations were explicitly political or partisan (e.g., ``Jair M. Bolsonaro,'' representing the page of
the far-right candidate for president of Brazil in 2018), indicating more
homogeneous audience reactions after accounting for the characteristics
included in the model. This contrast is broadly consistent with prior work
showing differences in interaction patterns between more neutral brokers and
partisan pages in the Brazilian political context
\cite{Kobellarz2024}. These page-level examples are, however, only illustrative. The random effects capture residual heterogeneity in reaction homogeneity and do not establish audience ideology, political disagreement, audience composition, or the processes that produce these differences.

Fig~\ref{fig:account_effects}, panel (b), shows a positive correlation
between the estimated page random intercepts and random slopes for
toxicity ($r=0.81$). In other words, pages with higher estimated baseline reaction homogeneity tended to have toxicity slopes above the global average, whereas pages with lower estimated baselines tended to have slopes below it. Note that a slope above the global average can still represent a negative association. This pattern describes page-level
heterogeneity in the fitted model. Because many pages had few observed posts, individual page estimates should be interpreted cautiously.

Social Identity Theory~\cite{tajfel2001integrative} offers one interpretation
of this correlation, in which hostile language works differently for
different political audiences. In an online experiment, like-minded
incivility led participants to write more uncivil comments themselves,
while disagreeable incivility induced anger and
aversion~\cite{gervais2015incivility}. On social media, posts about the
political out-group were shared about twice as often as posts about the
in-group~\cite{rathje2021outgroup}. If toxic posts on pages with homogeneous
reactions reach mostly like-minded users, toxicity there would be less
likely to split reactions. Our data do not show who reads these posts or
whether their toxic language targets an out-group, so we treat this
correlation as variation across pages in the toxicity association, not as
evidence of in-group cohesion.

\section*{Conclusion}
\label{secConclusion}

This study examined how ideological discrepancy between Facebook pages and
shared news links was associated with audience reactions and with the toxicity
of the accompanying post text in a dataset centered on Brazil's 2018
presidential election. Cross-cutting content is often expected to mitigate
polarization, yet encounters across ideological lines do not necessarily foster
more constructive engagement \cite{bail2018exposure,cetinkaya2025cross}. At the
level of individual sharing events, our results are consistent with that view.
Posts at the upper end of Bias Discrepancy drew less homogeneous reactions than
posts in the middle of the distribution and carried the most toxic accompanying
text. Departure from the ideological sharing repertoire was not the only
condition associated with mixed reactions, however. Reaction homogeneity was
also lower when the shared news link was very close to that repertoire, and
Content Reaction Positivity was lowest at the same end, although the data do
not establish the mechanism underlying this unexpected pattern.

Posts by right-leaning pages also generally had more homogeneous reactions than
posts by left-leaning pages. This asymmetry is compatible with prior work
reporting stronger cohesion within right-leaning communities in the Brazilian
context \cite{Kobellarz2019,kobellarz2022reaching}, but reaction homogeneity
does not itself measure audience cohesion or political agreement, and the
smaller number of right-leaning pages and links limits the generalizability of
this comparison.

For RQ2, Content Reaction Positivity, which reflects how the same link was
received in other posts, had the largest association with reaction homogeneity,
and omitting it reduced the marginal $R^2$ of the main model from $0.762$ to
$0.154$ (\nameref{app:r27_checks}). Greater Bias Discrepancy remained
associated with lower reaction homogeneity after accounting for this link-level
reception, page and content bias, ideological extremity, toxicity, content
topics, and activity controls. The association between toxicity and reaction
homogeneity also varied across pages. Pages with higher baseline reaction
homogeneity tended to have more positive toxicity slopes, and pages with lower
baseline reaction homogeneity tended to have more negative slopes. Work on
social identity and incivility offers a possible explanation
\cite{tajfel2001integrative,gervais2015incivility,rathje2021outgroup}, which
our data cannot test.

These results should be interpreted in light of several limitations. First, content bias scores were derived from Twitter retweet behavior
\cite{kobellarz2022reaching}, whereas Facebook posts and audience reactions
were observed on Facebook. This approach assumes that the ideological signal
derived from Twitter is informative about the news links when they circulate on Facebook, even though user populations and interaction dynamics may differ
across platforms. In addition, Facebook's reaction affordances may shape both
Content Reaction Positivity and the Reaction Homogeneity Index, so their
absolute levels should be interpreted in the context of the platform.

Second, page bias is inferred from the ideologically labeled links observed
in a page's sharing history. Bias Discrepancy therefore measures departure from the page's ideological sharing repertoire, rather than
independently established page ideology or the page's intent in sharing a
particular link. Even so, for pages with at least 15 distinct links, the inferred page bias agreed
with an independent identity-based classification of page orientation for
62 of the 66 classified pages (see \nameref{app:publisher_validation}). In addition, sensitivity analyses restricted pages to at least two or five distinct bias-scored links and recalculated page
bias after excluding all shares of the focal link by the page. The negative
association between Bias Discrepancy and reaction homogeneity persisted in each analysis (see \nameref{app:r27_checks}). These checks address short observed
histories and the focal link's contribution to the page-bias estimate, but
they do not establish independently held page ideology.

Third, right-leaning content and pages are underrepresented in the dataset,
requiring caution when generalizing comparisons involving this group. Fourth,
two features of the reaction measures constrain their interpretation. Content
Reaction Positivity is computed from other posts sharing the same link and
captures a content-level reaction pattern that overlaps with the content random
intercept. We therefore treat it as a contextual predictor, not as an
independent behavioral mechanism. The Reaction Homogeneity Index is
mathematically related to the focal post's negative-reaction share, so results
from these two outcomes should not be treated as independent evidence of the
same association.

Fifth, the temporal relationship between the Twitter and Facebook data imposes
additional limitations. Content bias scores were estimated from Twitter
retweet behavior between October 1 and December 11, 2018. Of the Facebook
posts in the analytical sample, 4,099 (89.4\%) fell within this window, 222 (4.8\%) preceded it, 20 (0.4\%) followed it later in 2018, and 243 (5.3\%)
were published from 2019 to 2022. Restricting the model to the 4,099 posts within that window preserved the negative association between Bias
Discrepancy and reaction homogeneity (see \nameref{app:r27_checks}).
Nevertheless, for posts outside this window, $b_c$ is a link-level
ideological proxy estimated from 2018 Twitter circulation rather than a
contemporaneous measure of the link's ideological reception. Because the
Facebook data were collected retrospectively, the sample may also be affected
by survivorship if deleted or otherwise inaccessible posts could not be
observed. CrowdTangle further limits the analysis to the public Facebook
content available through its collection system, excluding private interactions.

Finally, toxicity was measured using the Perspective API, a widely used tool
with known limitations for Portuguese-language content. Prior work documents
lower classification performance for Portuguese than for English
\cite{buzelin2024change}, as well as limitations in capturing conversational
context, sarcasm, and culturally specific forms of political expression
\cite{xenos2021context,muralikumar2023human}. Processing the posts directly in
Portuguese avoids translation-related distortions
\cite{bell2025translate,KobellarzWebMedia2022}, but the resulting toxicity
scores should nevertheless be interpreted as approximate indicators of hostile
language rather than exhaustive measures of political incivility
\cite{gervais2025incivility}.

Future research could extend this work using more context-aware and culturally
grounded measures of toxicity and by examining whether similar relationships
between ideological discrepancy, audience reactions, and accompanying language
emerge in other political contexts and on platforms with different interaction
affordances.

\section*{Ethical considerations}

The Facebook data analyzed in this study were collected from publicly accessible posts using the now-deprecated CrowdTangle API. The Twitter-derived content bias scores were obtained from the previously published dataset of \cite{kobellarz2022reaching}. Data collection was conducted prior to the tool's discontinuation, in compliance with the platform's Terms of Service and data use policies active at the time of access. The dataset comprises content exclusively from public pages (e.g., media outlets, political figures, and organizations) operating in the public sphere. While we report actor-specific heterogeneity to describe how distinct editorial lines or political stances relate to reaction homogeneity, this analysis does not expose private user data. No personally identifiable information from private citizens was analyzed individually. Toxicity scores were computed exclusively from the publicly available text published by these pages,
while audience-related metrics, such as the Reaction Homogeneity Index, were computed from aggregate reaction counts without identifying individual users.

\section*{Acknowledgments}

This research is partially supported by the SocialNet project (grant 2023/00148-0 of FAPESP). This research was also partially supported by the National Council for Scientific and Technological Development - CNPq (processes 441444/2023-7, 407520/2026-0, 403646/2026-0, and 444724/2024-9). This research is also part of the INCT TILD-IAR funded by CNPq (proc. 408490/2024-1). The funders had no role in study design, data collection and analysis, decision to publish, or preparation of the manuscript.

\bibliographystyle{plos2025}

\section*{Supporting Information}

\subsection*{S1 Appendix}
\label{app:cutoff}
\subsubsection*{Empirical assessment of the minimum-reaction cutoff}

This appendix presents the empirical analysis used to determine the minimum-reaction cutoff applied in the preprocessing stage.

\begin{figure}[ht]
\centering
\includegraphics[width=\linewidth]{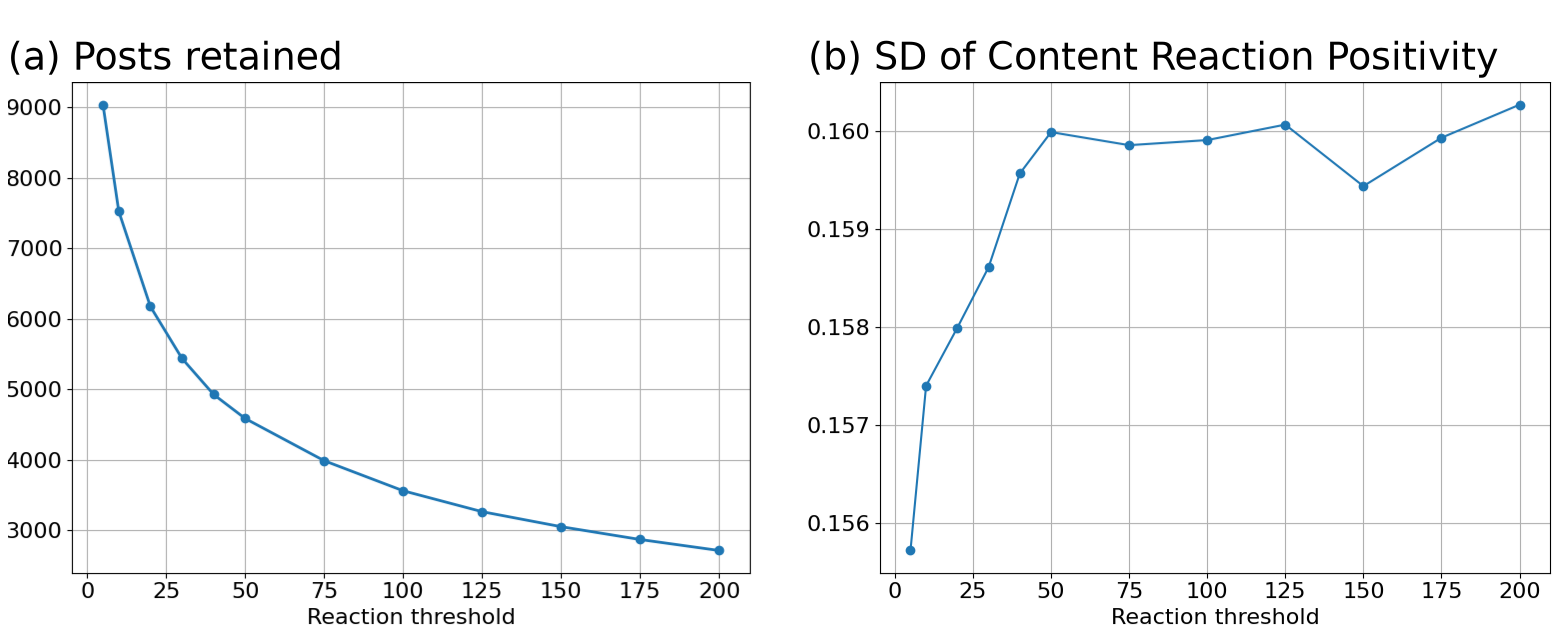}
\caption{Assessment of the minimum-reaction cutoff.
Panel (a) shows the number of retained posts at each cutoff. Panel (b)
shows the standard deviation of Content Reaction Positivity across the
retained posts. Dispersion rises toward a cutoff of 50 reactions and changes
relatively little thereafter. }
\label{fig:threshold_analysis}
\end{figure}

As shown in Fig~\ref{fig:threshold_analysis}, panel (a), increasing the minimum-reaction cutoff reduced the number of retained posts, with a less
steep decline at higher cutoffs. Panel (b) shows that the standard deviation of Content Reaction Positivity increased as the cutoff rose toward 50 reactions and changed relatively little at higher cutoffs. These curves
describe changes in sample size and between-post dispersion. They do not
directly measure the precision of individual scores. The 50-reaction cutoff
was an exploratory preprocessing choice. Sensitivity analyses using cutoffs
of 25 and 100 are reported in \nameref{app:r27_checks}.

\subsection*{S2 Appendix}
\label{app:topicSel}

\subsubsection*{Topic definition and model selection}

To characterize the thematic scope of the shared news articles, we first conducted an exploratory Latent Dirichlet Allocation (LDA) analysis \cite{blei2003latent} on the textual content of the linked news articles. The resulting latent themes informed the definition of ten topical categories used in the analysis: economy, education, health, security, culture, religion, disinformation, election, politics, and corruption. These predefined categories were then assigned to individual news articles using a zero-shot language model.

To select an appropriate zero-shot classification model, we conducted a comparative experiment with three candidate approaches. The first was the multilingual mDeBERTa-v3-base model fine-tuned on NLI data \cite{laurer2024less} (MoritzLaurer/mDeBERTa-v3-base-xnli-multilingual-nli-2mil7)\footnote{\url{https://huggingface.co/MoritzLaurer/mDeBERTa-v3-base-xnli-multilingual-nli-2mil7}}. The second was the BGE-M3 model \cite{laurer2023building} (MoritzLaurer/bge-m3-zeroshot-v2.0)\footnote{\url{https://huggingface.co/MoritzLaurer/bge-m3-zeroshot-v2.0}}. The third followed a translate-test pipeline using BART-Large-MNLI \cite{lewis-etal-2020-bart} (facebook/bart-large-mnli)\footnote{\url{https://huggingface.co/facebook/bart-large-mnli}}, where texts were first translated from Portuguese to English using the unicamp-dl/translation-pt-en-t5 model \cite{lopes-etal-2020-lite}.

For an exploratory model comparison, we manually labeled 50 news links for the presence of the 10 predefined topics. Of these, 48 matched the analytical news-link inputs used in the comparison, and the reported metrics were calculated on those 48 articles. Each model produced classifier scores $p_k \in [0,1]$ for topic $k$. Instead of applying a single global threshold, we optimized the classification threshold independently for each topic to account for differences in model sensitivity across domains. Binary predictions were defined as

\[
\widehat{y}_k =
\begin{cases}
1, & \text{if } p_k \geq \tau_k, \\
0, & \text{otherwise}.
\end{cases}
\]

For each topic $k$, the optimal threshold $\tau_k$ was selected by maximizing the F1-score on the validation set. We evaluated the F1-score at every unique probability value produced by the model, which avoids the discretization of a fixed-step grid search.

The thresholds and reported F1 scores were obtained from the same
48 labeled articles, without a separate test set. These scores therefore
summarize an exploratory comparison and may overestimate performance
on new articles. These thresholds were used only to compare the candidate
models. The main model used the continuous scores.

Model performance was summarized using the Weighted Average F1-score across topics. This aggregation accounts for class imbalance by weighting topic-level F1-scores according to their support (number of positive instances) in the validation set.

As summarized in Table~\ref{tab:model_selection}, the multilingual mDeBERTa-v3-base model achieved the best performance, with a Weighted F1-score of $0.825$, compared to $0.780$ for the translation-based BART-Large-MNLI pipeline and $0.768$ for BGE-M3. Based on this evaluation, we selected mDeBERTa-v3-base for topic classification in the main analysis.

\begin{table}[ht]
\centering
\footnotesize
\caption{Exploratory comparison of zero-shot topic classifiers using
48 matched manually labeled articles. Topic-specific thresholds were
selected and F1 scores calculated on the same articles.}
\label{tab:model_selection}
\begin{tabular}{lcc}
\hline
\textbf{Model Strategy} & \textbf{Approach} & \textbf{W-F1 Score} \\ \hline
\textbf{mDeBERTa-v3-base} & Multilingual & \textbf{0.825} \\
BART-Large-MNLI & Trans. (PT$\to$EN) & 0.780 \\
BGE-M3 & Multilingual & 0.768 \\ \hline
\end{tabular}
\end{table}

\subsection*{S3 Appendix}
\label{app:beta}

\subsubsection*{Choice of the beta regression framework}

The Reaction Homogeneity Index ($RHI$) is a continuous variable bounded in the unit interval $[0,1]$. Applying Ordinary Least Squares (OLS) regression to such data is problematic, as it violates key assumptions of homoscedasticity and normally distributed residuals. In bounded outcomes, variance depends on the mean and decreases as observations approach the boundaries of the support.

We therefore used a Beta regression framework within a Generalized Linear Mixed Model (GLMM). The Beta distribution is commonly used for proportions and rates. Its conditional variance depends on the mean, and its predicted values stay within the bounds of the outcome.

\subsubsection*{Boundary adjustment for the Reaction Homogeneity Index}

A limitation of the Beta distribution is that it is defined on the open interval $(0,1)$. However, the empirical Reaction Homogeneity Index included observations with values exactly equal to 0 or 1. Because the likelihood function of the Beta distribution is undefined at these boundaries, a transformation was required prior to estimation.

Following Smithson and Verkuilen~\cite{smithson2006better}, we applied a symmetric transformation that compresses the data into the open interval:

\begin{equation*}
RHI' = \frac{RHI (N - 1) + 0.5}{N},
\end{equation*}

where $RHI$ is the original Reaction Homogeneity Index and $N$ is the total sample size. The transformation is linear, so it preserves the ordering and shape of the distribution, and it keeps the likelihood defined for observations at 0 and 1.

\subsubsection*{Random effects structure}

The dataset had a hierarchical, cross-classified structure: individual pages generated multiple posts, and the same news links (content) were frequently shared across different pages. Ignoring this non-independence would lead to underestimated standard errors and inflated Type I error rates.

To account for this structure, we included random intercepts for both pages and content. We also included random slopes for $\text{Tox}$ at both levels, so that the association between $\text{Tox}$ and reaction homogeneity could differ across pages and news links instead of being assumed uniform.

\subsection*{S4 Appendix}
\label{app:publisher_validation}
\subsubsection*{Identity-based assessment of page bias}

We examined all 117 pages in the analytical sample with at least 15 distinct links in their sharing history. We assigned Left, Center, or Right labels using identifiable politicians, parties or movements, explicit ideological self-descriptions, and the editorial identity of recognizable media outlets. Pages without an identifiable orientation were left uncoded (51/117). These labels were assigned without reference to $b_p$. We then categorized $b_p$ as Left ($b_p\leq-0.3$), Center ($-0.3< b_p < 0.3$), or Right ($b_p\geq 0.3$). Table~\ref{tab:publisher_validation} compares the two classifications.

\begin{table}[ht]
\centering
\footnotesize
\caption{Identity-based labels compared with categories derived from $b_p$ (66 coded pages). Rows are identity-based labels and columns are $b_p$ categories.}
\label{tab:publisher_validation}
\begin{tabular}{lccc}
\hline
 & Left & Center & Right \\
\hline
Left   & 36 & 0 & 0 \\
Center & 1  & 8 & 2 \\
Right  & 0  & 1 & 18 \\
\hline
\end{tabular}
\end{table}

Agreement was 62/66 (93.9\%). One page whose name partially matched a different political movement was left uncoded after its identity was checked. Coding it by the name-matching rule alone would have yielded 62/67 agreement (92.5\%) and one Left--Right reversal. Media classifications require editorial judgments, and an audience-derived score for a media outlet need not match its editorial identity. This exercise therefore supports the face validity of $b_p$ for identifiable political pages but does not establish it as an independent measure of every page's ideology.

\subsection*{S5 Appendix}
\label{app:r27_checks}

\subsubsection*{Reaction composition and negative-reaction share}

Across the 4,584 analytical posts, Like accounted for 64.8\% of all
reactions. Considering the four reactions used to define positive and negative
reactions (Like, Love, Sad, and Angry), 953 posts (20.8\%) received only
positive reactions, while 3,631 (79.2\%) received reactions of both signs.
No post received only negative reactions. Only 237 posts (5.2\%) had a
majority of negative reactions, and eight posts had equal positive and
negative reaction counts. Let $f_i^- = n_i^-/(n_i^+ + n_i^-)$ denote the negative-reaction share of post $i$, the proportion of its positive- and negative-valence reactions that are negative. For the 4,347 posts (94.8\%) with $f_i^- \le 0.5$,

\[
RHI_i = 1 - 2f_i^-.
\]

The relationship between the negative-reaction share and $RHI$ is therefore
largely fixed by their definitions.

As an additional sensitivity analysis, we modeled the focal post's
negative-reaction share directly. Both negative-share models described below retained the main model's signed bias and extremity terms, activity controls,
topic predictors, and random-effects structure, while omitting $CRP$.
Greater Bias Discrepancy was associated with a higher negative-reaction
share in a mixed beta model
(standardized $\beta=0.147$, $p<0.001$). We also estimated a beta-binomial
model using the underlying negative and positive reaction counts. This model
likewise showed that greater Bias Discrepancy was associated with a larger
share of negative reactions ($\beta=0.190$, $p<0.001$).

Because the negative-reaction share and $RHI$ are mathematically related for
most posts, these models are a sensitivity analysis of the reaction pattern
captured by $RHI$, not independent evidence for a separate outcome.

\subsubsection*{Model specification and sensitivity analyses}

We assessed whether the association between Bias Discrepancy and reaction
homogeneity was sensitive to the representation of political bias and to
additional model specifications. Using the same posts, outcome, and remaining
covariates, the original specification containing signed content and page bias
had an AIC of $-14989.3$. Adding content and page extremity improved model fit
(AIC $=-15014.6$), and a likelihood-ratio test comparing this model with the
original signed-bias specification was significant
($\chi^2(2)=29.24$, $p<0.001$). Replacing the continuous content and page bias
terms with Left/Center/Right categories yielded an AIC of $-14998.8$.
Of these three alternatives, the specification containing both
signed bias and absolute extremity provided the best fit. The main model
reported in Table~\ref{tab:beta_results} additionally included total reactions and page links as activity controls (AIC $=-15046.2$).

We next assessed the sensitivity of the key associations using the main model as the reference.
Table~\ref{tab:robustness_final} reports standardized coefficients for Bias
Discrepancy and Toxicity after omitting Content Reaction Positivity,
restricting pages by the number of distinct bias-scored links in their sharing histories, restricting posts to the Twitter collection
window (October 1--December 11, 2018), reconstructing page bias without the
focal link, and varying the minimum-reaction cutoff.

For the sharing-history restrictions, we counted only distinct links with a content bias score, which are the links that contribute to $b_p$.
The page-activity control retained its original definition based on all
distinct shared links. In the leave-link-out analysis, we removed all posts
by a page sharing the focal news link from that page's sharing history,
then recalculated page bias, page extremity, and Bias Discrepancy together.
This excluded 371 posts from pages with no remaining bias-scored history,
leaving 4,213 posts. As in the original calculation, a remaining history with only center-leaning links was assigned $b_p=0$. Other predictors retained their
original definitions. The 25-reaction sample was rebuilt from the source
data using the same preprocessing and metric-construction procedures before applying the lower minimum-reaction cutoff. In every fitted sample, continuous
predictors were standardized and the boundary adjustment of $RHI$ was
recomputed using that sample's size.

\begin{table}[htb]
\centering
\footnotesize
\setlength{\tabcolsep}{3pt}
\caption{Sensitivity analyses of the main model for the Reaction Homogeneity Index. Except for the stated modification, each model
retains the fixed- and random-effects specification in
Table~\ref{tab:beta_results}. Continuous predictors are standardized within
each fitted sample. Coefficients are on the logit scale and intervals are
two-sided 95\% Wald confidence intervals.}
\label{tab:robustness_final}
\begin{tabular}{p{3.7cm}ccccc}
\toprule
\textbf{Specification}
& $\beta_{\Delta_b}$
& \textbf{95\% CI}
& $\beta_{\mathrm{Tox}}$
& \textbf{95\% CI}
& \textbf{N} \\
\midrule
Main model
& $-0.173$ & $[-0.211, -0.135]$
& $-0.149$ & $[-0.196, -0.102]$ & 4,584 \\

CRP omitted
& $-0.178$ & $[-0.219, -0.138]$
& $-0.133$ & $[-0.180, -0.086]$ & 4,584 \\

Pages with $\geq 2$ bias-scored links
& $-0.201$ & $[-0.241, -0.160]$
& $-0.162$ & $[-0.211, -0.113]$ & 4,213 \\

Pages with $\geq 5$ bias-scored links
& $-0.219$ & $[-0.264, -0.175]$
& $-0.170$ & $[-0.226, -0.113]$ & 3,405 \\

Twitter-window posts only
& $-0.173$ & $[-0.214, -0.132]$
& $-0.144$ & $[-0.192, -0.096]$ & 4,099 \\

Leave-link-out page bias
& $-0.192$ & $[-0.234, -0.150]$
& $-0.165$ & $[-0.214, -0.116]$ & 4,213 \\

Minimum 25 reactions
& $-0.171$ & $[-0.206, -0.135]$
& $-0.133$ & $[-0.179, -0.088]$ & 5,774 \\

Minimum 100 reactions
& $-0.167$ & $[-0.209, -0.126]$
& $-0.177$ & $[-0.229, -0.126]$ & 3,558 \\
\bottomrule
\end{tabular}
\end{table}

The main model and all seven sensitivity analyses returned an optimizer convergence code of zero
and a positive-definite Hessian, with finite coefficient estimates and
confidence intervals. Across all eight specifications, Bias Discrepancy
and Toxicity had negative coefficients, with all 95\% confidence intervals
excluding zero. The Bias Discrepancy association persisted after removing
the focal link from the page-bias calculation, after restricting the sample
to posts within the Twitter collection window, and under alternative minimum-reaction cutoffs. These checks support the robustness of the reported
association to these specific analytical choices. They do not remove the
limitations of estimating page ideology from observed sharing behavior or
establish a causal effect. Because standardization was performed within each
sample, coefficient magnitudes reflect the predictor distributions in
those samples.

Omitting $CRP$ from the main model reduced marginal $R^2$
from $0.762$ to $0.154$, while conditional $R^2$ remained similar
($0.977$ versus $0.979$). Marginal and conditional $R^2$ were calculated
using the Nakagawa method implemented in
\texttt{performance::r2\_nakagawa()}, with the lognormal approximation.
This indicates that $CRP$ accounted for substantial fixed-effect variation, while the Bias Discrepancy association persisted when this same-link reaction covariate was omitted.

\subsubsection*{Categorical bias specification}

As a sensitivity analysis, we replaced the signed content and page bias terms
with Left, Center, and Right categories, defined as $b\leq -0.3$,
$-0.3<b<0.3$, and $b\geq 0.3$, respectively.
Center was the reference category. The model used the same posts, outcome, and other predictors as the signed-bias model used for the specification
comparison, without the two activity controls.
Table~\ref{tab:categorical_bias} reports the bias terms and key associations.

\begin{table}[htb]
\centering
\footnotesize
\caption{Categorical bias specification for the Reaction Homogeneity Index.
Center is the reference category and continuous predictors are standardized.}
\label{tab:categorical_bias}
\begin{tabular}{lccc}
\toprule
Predictor & $\beta$ & 95\% CI & $p$ \\
\midrule
Content: Left
& $-0.047$ & $-0.169$ to $0.075$ & $0.451$ \\
Content: Right
& $0.244$ & $0.079$ to $0.408$ & $0.004$ \\
Page: Left
& $0.074$ & $-0.067$ to $0.215$ & $0.304$ \\
Page: Right
& $0.308$ & $0.154$ to $0.463$ & $<0.001$ \\
Bias Discrepancy ($\Delta_b$)
& $-0.163$ & $-0.201$ to $-0.125$ & $<0.001$ \\
Toxicity
& $-0.146$ & $-0.194$ to $-0.098$ & $<0.001$ \\
Content Reaction Positivity
& $0.962$ & $0.899$ to $1.024$ & $<0.001$ \\
\midrule
AIC & \multicolumn{3}{c}{$-14998.8$} \\
\bottomrule
\end{tabular}
\end{table}

The categorical specification yielded the same central conclusion regarding Bias Discrepancy: greater discrepancy remained associated with lower reaction homogeneity ($\beta=-0.163$, $p<0.001$). The categorical terms additionally showed higher $RHI$ for right-leaning content and right-leaning pages relative to their respective center categories, whereas the corresponding left-versus-center differences were not statistically significant.

\subsection*{S6 Appendix}
\label{app:discrepancy_distribution}
\subsubsection*{Composition of Bias Discrepancy quartiles}

Bias Discrepancy ($\Delta_b$) was divided into four groups using cut points of $0.161$, $0.313$, and $0.498$. The maximum observed value was $1.746$. The groups had nearly equal numbers of posts by construction. Table~\ref{tab:discrepancy_composition} reports the numbers of distinct pages and links and the shares contributed by the largest pages.

\begin{table}[htb]
\centering
\footnotesize
\caption{Composition and page concentration of Bias Discrepancy quartiles.}
\label{tab:discrepancy_composition}
\begin{tabular}{lrrrrr}
\toprule
Discrepancy level & Posts & Pages & Links & Largest page & Top 10 pages \\
\midrule
Very Low  & 1,148 & 427 & 361 & 4.4\% & 23.3\% \\
Low       & 1,148 & 554 & 419 & 6.7\% & 24.1\% \\
High      & 1,142 & 499 & 375 & 8.0\% & 27.0\% \\
Very High & 1,146 & 595 & 374 & 3.9\% & 19.1\% \\
\bottomrule
\end{tabular}
\end{table}

No group was dominated by a single page. The Very High discrepancy group had the most distinct pages and the smallest shares contributed by its largest page and top ten pages.

\subsection*{S7 Appendix}
\label{app:TwitterBias}
\subsubsection*{Estimation and validation of Twitter-derived content bias scores}

This appendix summarizes the procedure used by
\cite{kobellarz2022reaching} to estimate the Twitter-derived
content bias scores ($b_c$) employed in our study. In the original study, user polarity was estimated on
a weekly basis. Users with fewer than five tweets in
a given week were excluded, resulting in 72,576 users
in the Brazilian dataset. The hashtags involved were those present in each
user's tweets for that week, including the ones they
retweeted, rather than hashtags attached to the news links.

Reading a retweet as a signal of the retweeter's own
alignment is an established basis for estimating political
position from Twitter interaction data
\cite{conover2011political,conover2011predicting}, and it
is how retweeting was used throughout the original study
and in prior work on the same election
\cite{Kobellarz2019}.

Hashtags were labeled in two stages. The 100 most frequent
hashtags were classified as left, neutral, right, or
uncertain by three volunteers who were not authors of the
original study and who worked independently of one another.
Only the 64 hashtags on which all three agreed were retained
(Fleiss' $\kappa=0.63$). The remaining hashtags were labeled
by propagating these seeds through a co-occurrence network,
in which a node represents a hashtag and an edge joins two
hashtags used in the same tweet. This procedure classified
$57{,}487$ hashtags ($65.6\%$ of the $87{,}620$ observed).
Repeating the propagation 100 times while randomly hiding
$10\%$ of the manual labels gave $\kappa=0.84$, while
hiding $20\%$ gave $\kappa=0.78$.

Each user's polarity was then computed as the contrast
between the right- and left-leaning hashtags they used,
considering all their classified hashtags, with per-class
weights to account for the uneven numbers of left, neutral,
and right hashtags in the dataset. The score for a news link aggregated the polarity of the users who retweeted it.
Retweets were first counted across 21 user-polarity bins,
from $-1.0$ to $+1.0$ in increments of $0.1$. These counts
were normalized across the dataset and within each link
before being weighted by the corresponding polarity values
to obtain the content bias score. The resulting score is
interpreted on the same scale as user polarity
\cite{kobellarz2022reaching}. Thus, hashtags did not characterize the news links directly. Instead, they provided the initial ideological signal used to estimate user polarity, from which the content bias score was derived.

The original study also evaluated potential circularity
in this procedure using synthetic null models. The
correlations induced by the construction of the metric
were weak relative to those observed in the empirical
data, providing evidence that circularity had a limited
impact on the reported associations
\cite{kobellarz2022reaching}.

\nolinenumbers

\end{document}